\pdfoutput=1
\documentclass[journal]{IEEEtran}

\usepackage{lineno}
\usepackage{amsmath}
\usepackage{xcolor}
\usepackage{framed}
\usepackage{graphicx}
\usepackage{tikz}
\usepackage{color}
\usepackage{multirow}
\usetikzlibrary{shapes,arrows}
\usepackage{subcaption}
\usepackage{algorithm}
\usepackage[normalem]{ulem}
\usepackage[algo2e]{algorithm2e}

\newcommand{\norm}[1]{\left\lVert#1\right\rVert}
\newcommand{\Tr}[1]{\text{Tr}\left(#1\right) }
\newcommand{\vt}[1]{\text{vec}\left(#1\right) }
\newcommand*{\tran}{^{\mkern-1.5mu\mathsf{T}}}

\newcommand\blfootnote[1]{%
	\begingroup
	\renewcommand\thefootnote{}\footnote{#1}%
	\addtocounter{footnote}{-1}%
	\endgroup
}
\colorlet{shadecolor}{blue!20}

\usepackage[style=numeric, sorting=none, giveninits=true, backend=bibtex, maxbibnames=7, url=false, isbn=false, doi=false]{biblatex}
\bibliography{library}

\title{Tangent space spatial filters for interpretable and efficient Riemannian classification}
\author{
	\IEEEauthorblockN
	{
		Jiachen Xu\IEEEauthorrefmark{1},\textit{ Student Member, IEEE},
		Moritz Grosse-Wentrup,\textit{ Member, IEEE}
		and Vinay Jayaram,\textit{ Member, IEEE}
	}
}
\makeatletter
\def\blx@maxline{77}
\makeatother

\nolinenumbers
\begin{document}

\numberwithin{equation}{section}
\numberwithin{figure}{section}
\numberwithin{table}{section}

\maketitle

\blfootnote{Asterisk indicates corresponding author.}
\blfootnote{\IEEEauthorrefmark{1}J. Xu is with the Faculty of Computer Science, University of Vienna, H\"{o}rlgasse 6, 1090 Vienna, Austria (email: jxu1809@gmail.com).}
\blfootnote{M. Grosse-Wentrup is with the Faculty of Computer Science, University of Vienna, H\"{o}rlgasse 6, 1090 Vienna, Austria (email: moritzgw@ieee.org).}
\blfootnote{V. Jayaram is with the Department of Empirical Inference, Max Planck Institute for Intelligent Systems, Max-Planck-Ring 4, 72076 T\"{u}bingen, Germany (email: vjayaram@tuebingen.mpg.de).}

\begin{abstract}
	Methods based on Riemannian geometry have proven themselves to be good models for decoding in brain-computer interfacing (BCI). However, one major drawback of these methods is that it is not possible to determine what aspect of the signal the classifier is built on, leaving open the possibility that artifacts drive classification performance. In areas where artifactual control is problematic, specifically neurofeedback and BCIs in patient populations, this has led people to continue to rely on spatial filters as a way of generating features that are provably brain-related. Furthermore, these methods also suffer from the curse of dimensionality and are almost infeasible in high-density online BCI systems. To tackle these drawbacks, we introduce here a method for computing spatial filters from any linear function in the Riemannian tangent space, which allows for more efficient classification as well as the removal of artifact sources from classifiers built on Riemannian methods. We first prove a fundamental relationship between certain tangent spaces and spatial filtering methods, including an explanation of common spatial patterns within this framework, and then validate our proposed approach using an open-access BCI analysis framework.

\end{abstract}
\begin{IEEEkeywords}
	Brain-computer interface, Spatial filters, Riemannian geometry, Interpretability, Meta-analysis.
\end{IEEEkeywords}

\section{Introduction}
Brain-computer interfaces (BCIs) are well known, if not infamous, for their sensitivity to noise and their low signal-to-noise ratio. Over the past decades, many methods have been invented in order to derive features from the raw signal data that are predictive of user intention. However, as the electroencephalogram (EEG) is highly sensitive to both neural and non-neural signals, optimizing setups for predictive accuracy was insufficient. Rather, it was necessary to be able to confirm that any classifier was both predictive and based purely on brain-derived features. These divergent requirements spurred the field to develop in two different directions: spatial filtering and Riemannian manifold techniques.

Spatial filters are linear combinations of channel activity that reconstruct a single (neural or non-neural) source with certain desired properties. Initially, these weightings were computed via physical or neurophysiological models \cite{Nunez1991}. However, it was quickly discovered that data-driven spatial filters could lead to features that reflect robust differences in brain activity. By optimizing for variance \cite{grosse2009beamforming,subasi2010eeg} or independence \cite{Vigario2000,Makeig30091997}, or even searching for filters that maximize the difference between multiple types of intention \cite{koles1990spatial,ramoser2000optimal}, many different sorts of spatial filters can be computed. In order to verify that the reconstructed signal comes from the brain, it was possible to plot the spatial patterns corresponding to those filters on the scalp.

Beginning with common spatial patterns (CSP), there has been a large body of literature dedicated to finding algorithms that optimally reconstruct source activity based on a given criterion. For differences between two classes, CSP has proven itself to be robust and easy to implement (for a more exhaustive review, see \cite{Lotte2011}). More recently, methods have been developed to find sources that track a continuous variable of interest \cite{dahne2014spoc}. One major difficulty that was recognized early on is that, while it is relatively simple to generate appropriate spatial filters for data that is already recorded, the application of these filters to new data is often confounded by the highly non-stationary nature of the EEG signal. Filters that persist across multiple recording sessions, or filters that work on multiple subjects, are both open areas of research. Some groups look at different criteria to derive robust filters \cite{Martin-Clemente2018,Samek2013}, others use more probabilistic techniques \cite{Kang2011, WeiWu2009}, and still others consider options like sparsity \cite{Onaran2013,Goksu2011} or looking at patches of channels \cite{Sannelli2016}. Each of the aforementioned techniques has shown its value in solving an aspect of the spatial filtering problem, but they often require very different approaches to solve the ensuing optimization problems, and it is hard to decide which one may be most appropriate for a given situation. A further issue is the inefficient use of data. Spatial filters themselves can only be used to generate features. To fit classification or regression models requires re-using the data for the prediction model fitting.

Nevertheless, spatial filtering remains a crucial method in applications where artifacts are of great concern. In particular, it is crucial for neurofeedback studies. When the goal is to give feedback on neural activity, there must be a way of ensuring that the model which reconstructs a source of interest from the original signal only uses brain data to do so. This requirement invalidates many black-box machine learning methods, such as random forests \cite{fraiwan2012automated}, and often results in methodologies in which spatial filters are retrained in each recording session and validated by hand before neurofeedback can take place.

Outside of this sphere, methods based on Riemannian geometry have been gaining momentum as a model for robust classification for performance-optimized BCIs. Thanks to work in differential geometry, metrics for computing the distance between sensor covariance matrices have been discovered that are invariant to many common sorts of noise found in the electroencephalogram \cite{Moakher2005}. These methods can be translated into algorithms for finding classifiers that are far more robust to noise across a variety of contexts \cite{Barachant2010}. In particular, the approaches that use tangent space projection \cite{Barachant2013} have been shown to out-perform most other conventional methods in a recent meta-analysis \cite{jayaram2018moabb}. Two major downsides, however, are their high computational complexity and their interpretation. Because these methods work in the space of sensor covariance matrices, their size scales quadratically with the number of sensors. Further, the issue of interpretation is a significant problem. As of now, it is not possible to determine what parts of a signal are being used to build a tangent space classifier, and therefore these can only be used in artifact-sensitive contexts when paired with an artifact detection pipeline or other artifact cleaning methods.

In our paper, we show the following contributions: that it is possible to find sets of spatial filters that describe a linear function in the Riemannian tangent space, and further that this space has a fundamental relationship to common spatial patterns. Via this approach, the full literature of linear machine learning methods can be used for spatial filtering, instead of requiring a different optimization for each regularization of interest. Using this connection, it is possible to visualize the sources that a tangent space classifier uses, and thereby to identify artifact sources used for classification and remove them via orthogonal projection. Finally, we show in offline comparison that using spatial filters derived via this approach significantly out-performs common spatial patterns and can even, in low-data situations, out-perform the tangent space function they are derived from. A preliminary version of this work has been reported at \cite{Xuetal19}.

\section{Background}
Riemannian manifold-based classification methods (hereafter abbreviated Riemannian methods) can often seem difficult to understand. For convenience, we include this section that reviews our notation and the basic operations of Riemannian methods, as well as a short review of the mathematics behind spatial filtering.

\subsection{Preliminary and Notations} \label{notations}

We notate the the raw sensor data as $\mathbf{X}$ $\in$ $\Re^{C \times N \times T} $, where $C$, $N$ and $T$ represents the number of channels (electrodes), samples (length of each trial) and trials respectively. We represent the data of channel $c$ (with $c \in \{1, \cdots, C\}$) as $\mathbf{X}_c$. In addition, the data from the $t$-th trial (with $t \in \{1, \cdots, T\}$) are expressed as $\mathbf{X}^{t}$. Similarly, we use $(\cdot)^{t}$ to express the variables derived from $\mathbf{X}^{t}$. Moreover, the covariance matrices computed from $\mathbf{X}$, i.e., the points lying on the manifold, are denoted as $\mathbf{C}\in \Re^{C \times C \times T} $. The Fr\'{e}chet mean, a generalization of the standard arithmetic mean to other spaces, of the manifold points set $\mathbf{C}$ is expressed as  $\mathbf{C}^m$. In the following section, we use $\mathbf{A}$ to denote any symmetric positive definite (SPD) matrix, for which the following property holds true: $\mathbf{v}^T\mathbf{A}\mathbf{v} > 0, \forall \mathbf{v} \neq \mathbf{0}$. 

We next describe some common operations for manipulating points on the symmetric positive definite (SPD) manifold. Firstly, $\lambda\left( \mathbf{A} \right)$ is used to express the vector of eigenvalues of $\mathbf{A}$. Next, the logarithm for an SPD matrix is defined as: 
\begin{equation}
\label{Eq:2-1}
\begin{aligned}
\text{Logm}\left( \mathbf{A}\right) =  \mathbf{V} \text{log}\left( \mathbf{D}\right) \mathbf{V}\tran, 
\end{aligned}
\end{equation}
where $\mathbf{D}$ is the diagonal eigenvalue matrix of $\mathbf{A}$, i.e., $ \mathbf{A} = \mathbf{V}  \mathbf{D} \mathbf{V}^{T}$, $\text{log}\left( \cdot \right)$ represents taking the logarithm elementwise for a matrix, and $\mathbf{V}$ is the orthogonal matrix of eigenvectors. The exponential and powers of a SPD matrix are defined in similar fashion, i.e.:

\begin{equation}
\label{Eq:2-2}
\begin{aligned}
& \text{Expm}\left( \mathbf{A}\right) =  \mathbf{V} \text{exp}\left( \mathbf{D}\right) \mathbf{V}\tran \\
& \mathbf{A}^{p} =  \mathbf{V} \mathbf{D}^{p} \mathbf{V}\tran, \text{ }p \in \Re \text{ and } p \neq 0, 
\end{aligned}
\end{equation}
where $\text{exp}\left( \cdot \right)$ and $\left( \cdot \right)^{p}$ represent taking logarithm and power of $p$ elementwise within a matrix. Please note that $p$ can also be a fraction, e.g., $p=\frac{1}{2}$ means the square root and $p=-\frac{1}{2}$ denotes the inverse square root.

At last, since the vectorization of an SPD matrix is also frequently employed to reduce the computational complexity, it is defined as below:
\begin{equation}
\label{Eq:2-4}
\begin{aligned}
\vt{\mathbf{A}} =& \left[\alpha_{1,1}\mathbf{A}_{1,1}, \cdots, \alpha_{i,j}\mathbf{A}_{i,j}, \cdots,\alpha_{C,C}\mathbf{A}_{C,C} \right] \\ 
&\in \Re^{1\times \frac{C(C+1)}{2}}, \text{ where } 1 \leq j \leq i \leq C. \\
\alpha_{i,j} =& \left[ 1 \text{ if } i=j, \sqrt{2} \text{ else}  \right] 
\end{aligned}
\end{equation}
An overview of these notations is shown in Table \ref{T1}. 

\begin{table}[htb!]
    \caption{\protect \centering The List of Notations}%
    \label{T1}
    \centering
    \begin{tabular}{|l|l|}
        \hline
        \multicolumn{2}{|c|}{\textbf{Basic Variables}}  \\
        \hline
        $C - $ \text{  \#channels}              &   $c - $ \text{  $c$-th channel}                  \\
        $T - $ \text{  \#trials}                &   $t - $ \text{  t-th trial}                    \\
        $N - $ \text{  \#samples}               &   $K - $ \text{  \#spatial filters}             \\
        $\mathbf{D} - \text{Diagonal matrix}$  &   $\mathbf{I} - \text{Identity matrix}$        \\
        $\mathbf{A} - \text{Any SPD matrix}$   &   $\mathbf{B} - \text{Any SPD matrix}$  \\                                                                                                        
        \hline
        \multicolumn{2}{|c|}{\textbf{Data Related Variables}} \\    
        \hline
        \multicolumn{2}{|l|}{$\mathbf{X} \in \Re^{C \times N \times T}  - \text{Bandpass filtered trialwise data} $} \\
        \multicolumn{2}{|l|}{$\mathbf{C}\in \Re^{C \times C \times T}   - \text{Covariance matrices on the manifold}$} \\    
        \multicolumn{2}{|l|}{$\mathbf{C}^w  \in \Re^{C \times C} - \text{Weight covariance matrix on the manifold} $} \\
        \multicolumn{2}{|l|}{$\mathbf{C}_{\text{ref}}  \in \Re^{C \times C} - \text{Reference point for constructing tangent space} $} \\    
        \multicolumn{2}{|l|}{$\mathbf{S} \in \Re^{C \times  C \times T}   - \text{Covariance matrices on the tangent space}$ } \\    
        \multicolumn{2}{|l|}{$\mathbf{S}^w  \in \Re^{C \times C} - \text{Weight matrix on the tangent space} $} \\    
        \multicolumn{2}{|l|}{$\mathbf{F} \in \Re^{C \times C} - \text{Spatial filters with full rank}$} \\    
        \multicolumn{2}{|l|}{$\overrightarrow{s^t} \in \Re^{\frac{C(C+1)}{2}\times 1} - \text{Tangent vector of $t$-th trial}$} \\
        \multicolumn{2}{|l|}{$\overrightarrow{f} \in \Re^{C \times 1} - \text{Single spatial filter component} $} \\    
        \multicolumn{2}{|l|}{$\overrightarrow{w} \in \Re^{\frac{C(C+1)}{2} \times 1} - \text{Weight vector on the tangent space} $} \\
        \multicolumn{2}{|l|}{$\overrightarrow{\beta} \in \Re^{C \times 1} - \text{Vector of sorted regression coefficient (log-eigenvalue)} $} \\
        \hline
        \multicolumn{2}{|c|}{\textbf{Operators}} \\    
        \hline
        \multicolumn{2}{|l|}{$(\cdot)^{t} - \text{Variables from the data of $t$-th trial}$} \\
        \multicolumn{2}{|l|}{$\widetilde{(\cdot)}_{\perp \mathbf{F}} - \text{After spatial filtering with }\mathbf{F} $} \\
        \multicolumn{2}{|l|}{$(\cdot)^m - \text{Fr\'{e}chet mean}$} \\
        \multicolumn{2}{|l|}{$ \overline{(\cdot)}  - \text{Arithmetic mean}$} \\
        \multicolumn{2}{|l|}{$\text{vec}\left( \cdot \right) - \text{  Vectorizing SPD matrices}$} \\
        \multicolumn{2}{|l|}{$\lambda\left( \cdot \right) - \text{  The eigenvalue vector of a matrix }$ } \\
        \multicolumn{2}{|l|}{$\text{log}\left( \cdot \right) - \text{Taking logrithm elementwise}$} \\
        \multicolumn{2}{|l|}{$\text{Logm}\left( \cdot \right) - \text{Taking logrithm for a matrix based on the } \mathbf{I} $ } \\
        \multicolumn{2}{|l|}{$\text{Expm}\left( \cdot \right) - \text{Taking exponential for a matrix based on the } \mathbf{I} $} \\
        \multicolumn{2}{|l|}{$\text{Logm}_{\mathbf{A}}\left( \cdot \right) -\text{Taking logrithm for a matrix based on the } \mathbf{A}  $} \\
        \multicolumn{2}{|l|}{$\text{Expm}_{\mathbf{A}}\left( \cdot \right) - \text{Taking exponential for a matrix based on the } \mathbf{A} $} \\
        \hline
    \end{tabular}
\end{table}

\subsection{Riemannian Manifold based Methods}

Most BCIs use classifiers built on features extracted from the bandpower in physiologically relevant ranges from the recorded channels, often approximated via the variance after spectral filtering. One limitation of this information is that it cannot take correlations between channels into account; in order to overcome this,  the Riemannian classification framework is based on the sample covariance matrices \cite{Barachant2010}, which encode both cross-channel information and variance information, and are also theoretically symmetric positive definite. Therefore, in this section, we will introduce several fundamental procedures utilized in Riemannian methods. For a more mathematically exhaustive treatment of Riemannian manifolds, please refer to \cite{boothby1986introduction}.

\subsubsection{Riemannian Metric and Distance}
Most feature extraction or classification algorithms concentrate on maximally increasing the discriminability of data points. A linear classifier, for example, can be thought of as a one-dimensional projection of a dataset in which the two classes are as far from each other as possible according to a given criterion. One convenient proxy of measuring the discriminability of a set of points is therefore via the inter-point distances. A classifiable dataset corresponds to a dataset in which inter-point distances are low within a class and high between classes. Therefore, a good metric can lead to good models in machine learning tasks. In standard vector algebra, the metric function is usually the standard Euclidean metric, i.e., the squared Euclidean distance between two matrices, and is usually measured by the Frobenius norm of their difference, as shown in the below:

\begin{equation}
\label{Eq:2-5}
\begin{aligned}
d_{\text{Euclid}}^2\left( \mathbf{A},  \mathbf{B}\right) = \norm{\mathbf{A} - \mathbf{B}}_{\text{F}}^2 = \Tr{\left(\mathbf{A} - \mathbf{B} \right)^2 },
\end{aligned}
\end{equation}
where $\mathbf{A} $ and $\mathbf{B} $ are two matrices of the same size and $\norm{\cdot}_{\text{F}}$ represents the Frobenius norm.

While this metric can also be used with SPD matrices,  it is incapable of adequately capturing the structure of SPD matrices, leading to certain undesirable effects such as the swelling effect \cite{arsigny2006log}. What this means is that naively attempting to use covariance matrices as features in a linear classifier by simply vectorizing the input points often works very poorly. 

In order to take advantage of the structure inherent to covariance matrices, it is desirable to have a metric that generalizes the properties of the Euclidean metric in standard vector spaces to the SPD manifold. One importance of such property is the idea of geodesic distances -- that the distance between two points is equivalent to the length of the shortest path to get from point A to point B. In vector spaces, this is simply equivalent to the magnitude of the difference between two points, but this is not necessarily true for manifolds. The affine-invariant Riemannian metric is proposed \cite{pennec2006riemannian} and defined as Eq.~(\ref{Eq:2-6}), and has the property of preserving geodesic distances, which is to say that the distance between two points is the length of the shortest path between them upon the SPD manifold. 

\begin{equation}
\label{Eq:2-6}
\begin{aligned}
d_{\text{AIRM}}^2\left( \mathbf{A},  \mathbf{B} \right) =  \norm{ \log\left( \lambda \left(  \mathbf{A}^{-\frac{1}{2}}  \mathbf{B}  \mathbf{A}^{-\frac{1}{2}} \right) \right)  }_2^2, 
\end{aligned}
\end{equation}
where $\norm{\cdot}_2$ represents the L2 norm.

Based on the chosen metric, the expression for the mean of a set of matrices is defined as: 
\begin{equation}
\label{Eq:2-7}
\begin{aligned}
\mathbf{C}^m =\underset{\mathbf{A} \in \mathbf{C}  }{\arg \min} \sum_{t=1}^{T} d_{\text{AIRM}}^2\left( \mathbf{A},  \mathbf{C}^t \right),
\end{aligned}
\end{equation}
where $\mathbf{A} \in \Re^{C \times C}$ and $\mathbf{C} = \left[ \mathbf{C}^1, \cdots, \mathbf{C}^T\right] \in \Re^{C \times C \times T}$.

If $\mathbf{C}^m$ is globally unique, then it is named as the Fr\'{e}chet mean of the set of SPD matrices $\mathbf{C}$. Given the metric and a set of points, it is possible to implement purely distance-based classifiers such as k-Nearest Neighbors or Minimum Distance to Riemannian Mean (MDRM) \cite{barachant2011multiclass}. Classifiers based on this metric have shown themselves to be highly effective in particular for BCI data \cite{Barachant2010, Congedo2017, Yger2017}.

\subsubsection{Tangent Space}
When observing the explicit Riemannian metric defined in Eq.~(\ref{Eq:2-6}), one inconvenient issue is that the length of the shortest path between two manifold points (the geodesic) cannot be derived via simple subtraction and norm computation, as it can with the Euclidean metric. In order to treat SPD matrices in a manner identical to traditional feature vectors, we adopt the tangent space mapping.

The tangent space is defined as a finite-dimensional Euclidean space which exists at each point on the manifold and linearizes the curvature of the manifold around that point, which is called the reference point. Simply speaking, it is a way of transforming manifold points such that we can now treat them as standard vectors. However, the distances and angles derived from the tangent space representation of points are only valid within a small neighborhood around the reference point, which means that this transformation only works in a small volume of the manifold. Therefore, to ensure the approximation error is minimized, the Fr\'{e}chet mean of a set $\mathbf{C}$ is adopted as the reference point for that set.

To transform points from the manifold to the tangent space at a point and vice versa, the so-called logarithmic and exponential maps are used. Under the affine-invariant Riemannian metric, the logarithm and exponential function pair at a point $\mathbf{A}$ are formulated as following: 
\begin{equation}
\label{Eq:2-8}
\begin{aligned}
&\mathbf{S}^t= \text{Logm}_{\mathbf{A}}\left( \mathbf{C}^t\right)  \\
&\mathbf{C}^t= \text{Expm}_{\mathbf{A}}\left( \mathbf{S}^t\right), 
\end{aligned}
\end{equation}
where $\mathbf{S}^t$ is the projected point lying on the tangent space, $\mathbf{C}^t$ is the original manifold point and the operation of $\text{Logm}_{\mathbf{B}}\left( \mathbf{A}\right)$ and $\text{Expm}_{\mathbf{B}}\left( \mathbf{A}\right) $, i.e., the logarithm and exponential of $ \mathbf{A}$ based on another SPD matrix $ \mathbf{B}$,  is defined as \cite{pennec2006riemannian, Barachant2010}:
\begin{equation}
\label{Eq:2-3}
\begin{aligned}
&\text{Logm}_{\mathbf{B}}\left( \mathbf{A}\right) =  \mathbf{B}^{\frac{1}{2}} \text{Logm}\left( \mathbf{B}^{-\frac{1}{2}}  \mathbf{A}  \mathbf{B}^{-\frac{1}{2}}  \right) \mathbf{B}^{\frac{1}{2}} \\
&\text{Expm}_{\mathbf{B}}\left( \mathbf{A}\right) =  \mathbf{B}^{\frac{1}{2}} \text{Expm}\left( \mathbf{B}^{-\frac{1}{2}}  \mathbf{A}  \mathbf{B}^{-\frac{1}{2}}  \right) \mathbf{B}^{\frac{1}{2}} \\
\end{aligned}
\end{equation} 

To further simplify operations on the tangent space, the projected points are usually vectorized. Note that this procedure does not alter the location or norm of the points, it simply makes them easier to notate and use. We denote these vectors as tangent vectors and formulate them as follows:
\begin{equation}
\label{Eq:2-9}
\begin{aligned}
&\overrightarrow{s^t} = \vt{\mathbf{S}^{t}} \in \Re^{ \frac{C(C+1)}{2} \times 1} , 
\end{aligned}
\end{equation}
where $\overrightarrow{s_t}$ is the tangent vectors of $t$-th trial.

After obtaining the set of tangent vectors $\mathbf{s}$, standard machine learning algorithms can be applied.

\subsubsection{Pros and Cons of Riemannian Methods}
As an emerging technique, Riemannian methods have seen an upsurge of interest in the BCI field recently \cite{Congedo2017, Yger2017} due to their rich feature space and robustness to outliers. In particular, Jayaram \textit{et al.} \cite{jayaram2018moabb} have compared Riemannian methods and standard processing pipelines over more than 200 subjects and showed that Riemannian methods are, on average, superior to many other conventional methods.

One major pitfall of these methods, however, is their sensitivity to the number of channels. As shown in Eq.~(\ref{Eq:2-9}), the dimension of the tangent vectors increases quadratically with the number of channels $C$. In addition, the computational complexity of the eigenvalue decomposition for matrices grows cubically. Due to these reasons, it becomes infeasible to apply Riemannian methods on data sets with a large number of channels. In addition, since the full covariance matrix is utilized for classification, interpreting the contribution from each channel that is used by the classifier can be a challenge. Therefore, the application of Riemannian methods is still restricted to low-channel situations where interpretability is of lesser importance.

\subsection{Spatial Filtering} \label{SF}

The novelty of Riemannian methods is not only the adoption of the new metric function but also the extraction of covariance matrix based features instead of variance-based features. Although in traditional EEG-based BCI systems, power (variance)-based methods are much more commonly adopted, they are, unfortunately, significantly affected by poor signal quality. To remove artifacts and noise while reducing the computational complexity, spatial filtering techniques are often used. Since the projection of the underlying neuronal sources to the EEG electrodes can be modeled as a linear transformation \cite{haufe2014interpretation}, with the appropriate projection, it is possible to recover the activity of specific parts of the brain. This both increases signal quality and provides a convenient signal for neuro-feedback. 

Based on the way the filters are extracted, they can be categorized into fixed weight and data-driven \cite{lotte2014tutorial}. Among the latter, one of the most popular methods is Common Spatial Patterns (CSP) \cite{koles1990spatial, ramoser2000optimal, lotte2007review}, which has had a great impact on BCIs in the past two decades. In addition to increase signal quality, spatial filters are also important in ensuring that features used for machine learning in a BCI are brain-based. Since spatial filters represent a decoding model of brain activity, in which the time-series of interest is distilled from the recorded data, it is possible under some assumptions to recover an encoding model, which shows how the desired source projects to the sensors. This projection, called a spatial pattern, can then be visually validated to confirm that it represents a current source within the brain. While patterns are typically recovered by inverting the filtering matrix, this is only exact when the spatial filtering matrix is full rank. Therefore, for the computation of spatial patterns, we adopted the method proposed in \cite{haufe2014interpretation}, which is a more general way to derive the spatial patterns from linear filters.

We next briefly review the mathematics behind CSP, as it is one of the most common and simple methods for generating spatial filters. As formulated below, CSP aims at extracting the signal sources which maximize the variance ratio between two conditions:
\begin{equation}
\label{Eq:2-10}
\begin{aligned}
\overrightarrow{f}_{\text{CSP}} =\underset{\overrightarrow{f} \in \Re^{C \times 1}   }{\arg \max} \frac{\overrightarrow{f}\tran\overline{\mathbf{C}^{(+)}}\overrightarrow{f}}{\overrightarrow{f}\tran\overline{\mathbf{C}^{(-)}}\overrightarrow{f}},
\end{aligned}
\end{equation}
where $\overrightarrow{f}_{\text{CSP}} $ represents the optimal CSP filter component, and  $\overline{\mathbf{C}^{(+)}}$ and $\overline{\mathbf{C}^{(-)}}$ represents the arithmetic mean covariance matrix of each condition. Obviously, Eq.~(\ref{Eq:2-10}) can be solved by the Generalized Eigenvalue Decomposition (GED) between $\overline{\mathbf{C}^{(+)}}$ and $\overline{\mathbf{C}^{(-)}}$. The spatial filter matrix with shape of $\Re^{C \times K}$ is extracted by selecting the eigenvetors correpsonding to the first $K$ largest GED eigenvalues.

While CSP is usually extracted by solving $\text{GED} (\overline{\mathbf{C}^{(+)}}, \overline{\mathbf{C}^{(-)}} )$, it can also be interepreted in a discriminative view, as described by Blankertz \textit{et al}~\cite{Blankertz2008}:

\begin{equation}
\label{Eq:2-26}
\begin{aligned}
\mathbf{C}_d &= \overline{\mathbf{C}^{(+)}} -  \overline{\mathbf{C}^{(-)}} \text{: discriminative activity} \\
\mathbf{C}_c &= \overline{\mathbf{C}^{(+)}} + \overline{\mathbf{C}^{(-)}}  \text{: common activity} \\
\end{aligned}
\end{equation}

Here the solution of CSP is obtained by maximizing the variance ratios between discriminative and common activity:
\begin{equation}
\label{Eq:2-27}
\begin{aligned}
\overrightarrow{f}_{\text{CSP}} =\underset{\overrightarrow{f} \in \Re^{C \times 1}   }{\arg \max} \frac{\overrightarrow{f}\tran\mathbf{C}_d\overrightarrow{f}}{\overrightarrow{f}\tran\mathbf{C}_c\overrightarrow{f}}
\end{aligned}
\end{equation}
Thus, the spatial filter matrix of CSP, i.e.,
\begin{equation}
\label{Eq:2-27-add}
\begin{aligned}
\mathbf{F}_{\text{CSP}} = \left[\overrightarrow{f}_{\text{CSP}, 1}, \cdots, \overrightarrow{f}_{\text{CSP}, c}, \cdots, \overrightarrow{f}_{\text{CSP}, C} \right] \in \Re^{C \times C}, 
\end{aligned}
\end{equation}
can be extracted via GED($\mathbf{C}_d$,$\mathbf{C}_c$), where $\overrightarrow{f}_{\text{CSP}, c}$ is the $c$-th spatial filter component of the matrix.

\section{Methods}
Utilizing the least dimension to achieve the highest discriminability is always the ideal when designing a feature extraction algorithm. Although the features extracted from standard Riemannian methods are of high quality, they are hamstrung by the curse of dimensionality and a lack of interpretability. It is striking that, when reviewing these two factors which impede the application of Riemannian methods, spatial filtering techniques seem to be the remedy. The arguments are two-fold: First, reducing the dimensionality of the covariance matrices decreases computation time drastically. Second, thanks to the associated spatial patterns of spatial filtering, it is possible to verify what aspects of the recorded signal are being used by the classifier. Hence, how to leverage the spatial filtering technique in the standard Riemannian methods becomes an interesting question.

Inspired by this idea, in this section, we first propose a novel spatial filter extraction algorithm in which we approximate a linear function on the Riemannian tangent space points by a set of spatial filters, which render that function much less computationally intensive and also more understandable. We support the proposed algorithm by rigorous mathematical proofs. Moreover, by adopting this approximation idea, a simplified regression-like classification method is also put forward. Subsequently, CSP is proven to be a special case of the proposed tangent space spatial filtering. We validate our theoretical findings experimentally via the validation setup proposed in \cite{jayaram2018moabb}.

Mathematically-oriented readers are invited to begin below; readers more interested in a practical understanding may refer to Section \ref{sec:summary}.

\subsection{The Approximation of Standard Riemannian Methods via Spatial Filtering} \label{approx_def}
For the tangent space based Riemannian methods, the decision function (or decision boundary) on the tangent space completely determines the classification accuracy, because the predicted labels are entirely based on the output of decision function. In order to unify spatial filtering and tangent space-based methods, one option is to attempt to find filters that can preserve this function. For simplicity, we consider linear functions in the tangent space:
\begin{equation}
\label{Eq:2-11}
\begin{aligned}
\hat{y}_t =\overrightarrow{w} \tran \overrightarrow{s^t}\in \Re^{1 \times 1}, \hskip 0.1cm \forall t = 1, ..., T, 
\end{aligned}
\end{equation}
where $\overrightarrow{w}$ is the weight vector on the tangent space and $\hat{y}_t $ represents the predicted label from the decision function for $t$-th trial. One thing that should be noted is that as a constant value, the bias term can be ignored in the above equation since the later proof of equivalence still hold if adding same bias to both sides. Moreover, based on the definition and property of AIRM \cite{pennec2006riemannian}, the inner product on the tangent space can be expressed as the function of manifold points as derived below:
\begin{equation}
\label{Eq:2-12}
\begin{aligned}
\hat{y}_t &= \overrightarrow{w} \tran \overrightarrow{s^t} \\
&=<\mathbf{S}^w, \mathbf{S}^t>\mid_{\mathbf{C}^m} \\ 
&= <\text{Logm}_{\mathbf{C}^m}(\mathbf{C}^w), \text{Logm}_{\mathbf{C}^m}(\mathbf{C}^t) >\mid_{\mathbf{C}^m}, \hskip 0.1cm \forall t = 1, ..., T, 
\end{aligned}
\end{equation}
where $\mathbf{S}^w$ is defined as the weight covariance matrix generated via the reshaping of the tangent space weight vector $\overrightarrow{w}$ into a symmetric matrix, and $\mathbf{C}^w$ is the weight covariance matrix re-projected onto the manifold via the exponential map $\text{Expm}_{\mathbf{C}^m}\left( \mathbf{S}^w \right) $. In addition, we use $(\cdot)\mid_{\mathbf{C}_{\text{ref}}}$ to represent the variable lying on the tangent space which is computed based on the reference point $\mathbf{C}_{\text{ref}}$ and the operation $<\mathbf{S}_1, \mathbf{S}_2>\mid_{\mathbf{C}_{\text{ref}}} $ is defined in below lemma:

\textbf{\textit{Lemma 1: Inner products between tangent vectors are invariant to affine transformation}}
\begin{equation}
\begin{aligned}
\label{Eq:Lemma-0-maintext}
<\mathbf{S}_1, \mathbf{S}_2>\mid_{\mathbf{C}_{\text{ref}}} &=<\mathbf{C}_{\text{ref}}^{-\frac{1}{2}}\mathbf{S}_1\mathbf{C}_{\text{ref}}^{-\frac{1}{2}}, \mathbf{C}_{\text{ref}}^{-\frac{1}{2}}\mathbf{S}_2\mathbf{C}_{\text{ref}}^{-\frac{1}{2}}>\mid_{\mathbf{I}}\\
&=\Tr{\mathbf{C}_{\text{ref}}^{-\frac{1}{2}}\mathbf{S}_1\mathbf{C}_{\text{ref}}^{-\frac{1}{2}} \cdot \mathbf{C}_{\text{ref}}^{-\frac{1}{2}}\mathbf{S}_2\mathbf{C}_{\text{ref}}^{-\frac{1}{2}}}, \\
\end{aligned}
\end{equation}
where $\mathbf{S}_1$, $\mathbf{S}_2$ are two matrix-formatted tangent vectors on the tangent space computed at reference point $\mathbf{C}_{\text{ref}}$. For a full proof please refer to Appendix {\ref{GED_approx_lemma1}}.

Similarly, the approximated predicted labels from all the manifold points which are passed through a spatial filtering matrix $\mathbf{F}$ are as expressed below:
\begin{equation}
\label{Eq:2-13}
\begin{aligned}
y_t^{\text{approx}}\mid_{\mathbf{F}}  &=  <\widetilde{\mathbf{S}}^w_{\perp \mathbf{F}}, \widetilde{\mathbf{S}}^t_{\perp \mathbf{F}}  >\mid_{\widetilde{\mathbf{C}}^m_{\perp \mathbf{F}}}  \\
&=  <  \text{Logm}_{\widetilde{\mathbf{C}}^{m}_{\perp \mathbf{F}}  }    (\widetilde{\mathbf{C}}^w_{\perp \mathbf{F}}),  \text{Logm}_{\widetilde{\mathbf{C}}^{m}_{\perp \mathbf{F}}}(\widetilde{\mathbf{C}}^t_{\perp \mathbf{F}})>\mid_{\widetilde{\mathbf{C}}^m_{\perp \mathbf{F}}}  \\
&=  <  \text{Logm}_{\mathbf{F}\tran\mathbf{C}^m\mathbf{F}}    (\mathbf{F}\tran\mathbf{C}^w\mathbf{F}),  \\ &\text{\hspace{0.75cm}}\text{Logm}_{\mathbf{F}\tran\mathbf{C}^m\mathbf{F}}(\mathbf{F}\tran\mathbf{C}^t\mathbf{F})>\mid_{\mathbf{F}\tran\mathbf{C}^m\mathbf{F}},  \\
\end{aligned}
\end{equation}
where $y_t^{\text{approx}}\mid_{\mathbf{F}}$ is denoted as $y_t^{\text{approx}}$ thereafter for the convenience of notation and $\widetilde{(\cdot)}_{\perp \mathbf{F}} $ represents the matrix after filtering, e.g., $\widetilde{\mathbf{A}}_{\perp \mathbf{F}} = \mathbf{F}\tran\mathbf{A}\mathbf{F}$. Note that a property of the AIRM is that, for full-rank filtering matrices $\mathbf{F}$, the Fr\'{e}chet mean of the filtered matrices is the filtered mean of the original matrices, i.e. $\widetilde{\mathbf{C}}^{m}_{\perp \mathbf{F}}= \mathbf{F}\tran\mathbf{C}^m\mathbf{F} $. 

After explicitly formulating the true and approximated decision function, the next problem remained to resolve is the extraction of the spatial filter matrix $\mathbf{F}$. The optimal scenario from the perspective of consequence is that this spatial filter matrix $\mathbf{F}$ can perfectly reconstruct the decision function. Hence, in the next subsection, we provide mathematically rigorous derivation and proof to find the optimal solution of $\mathbf{F}$.

\subsection{Optimal Spatial Filter Extraction from the Tangent Space} \label{sec_TSSF}
Naively, the goal of spatial filter extraction is to find a filtering matrix that maximally reconstructs the tangent space function, which is shown as follows:
\begin{equation}
\label{Eq:2-14}
\begin{aligned}
\mathbf{F}_K^{\ast} = \underset{ \mathbf{F}_K \in \Re^{N \times K}}{\arg \min} \sum^T(\hat{y}_t- y_t^{\text{approx}}\mid_{\mathbf{F}_K} )^2,
\end{aligned}
\end{equation}
where $\mathbf{F}_K^{\ast}$ is the optimal filter matrix composed of $K$ spatial filter components from full filter matrix $\mathbf{F}$.

After substituting $\hat{y}_t$ (Eq.~(\ref{Eq:2-12})) and $y_t^{\text{approx}}$ (Eq.~(\ref{Eq:2-13})) into Eq.~(\ref{Eq:2-14}), the objective function of the optimization becomes rather complicated and intractable, even if the cost function in Eq.~(\ref{Eq:2-14}) is only the squared loss.

Clearly, the major obstacle for solving this optimization problem lies in the complex formulation of $y_t^{\text{approx}}\mid_{\mathbf{F}_K} $. Considering that $\mathbf{F}_K$ is a subset of $\mathbf{F}$, we first focus on the structure of $y_t^{\text{approx}}$ to see whether it can be simplified in the case that $\mathbf{F}_K$ is full rank.

By leveraging \textit{lemma 1}, $y_t^{\text{approx}}$ is solved by substituting 
\begin{equation}
\label{Eq:Lemma-0-subs}
\begin{aligned}
\mathbf{S}_1 &= \text{Logm}_{\mathbf{F}\tran\mathbf{C}^m\mathbf{F}}    (\mathbf{F}\tran\mathbf{C}^w\mathbf{F}) \\ 
\mathbf{S}_2 &= \text{Logm}_{\mathbf{F}\tran\mathbf{C}^m\mathbf{F}}    (\mathbf{F}\tran\mathbf{C}^t\mathbf{F}), \forall t \in \left[1, 2, \cdots, T \right] \\ 
\mathbf{C}_{\text{ref}} &= \mathbf{F}\tran\mathbf{C}^m\mathbf{F}
\end{aligned}
\end{equation}
into \textit{lemma 1} and we can obtain:
\begin{equation}
\begin{aligned}
\label{Eq:Lemma-0-full_substitution}
y_t^{\text{approx}} &=  <  \text{Logm}_{\mathbf{F}\tran\mathbf{C}^m\mathbf{F}}    (\mathbf{F}\tran\mathbf{C}^w\mathbf{F}), \text{Logm}_{\mathbf{F}\tran\mathbf{C}^m\mathbf{F}}(\mathbf{F}\tran\mathbf{C}^t\mathbf{F})>\mid_{\mathbf{F}\tran\mathbf{C}^m\mathbf{F}}  \\
&= \text{Tr} \left( \left(  \mathbf{F}\tran\mathbf{C}^m\mathbf{F} \right) ^{-\frac{1}{2}}\text{Logm}_{\mathbf{F}\tran\mathbf{C}^m\mathbf{F}}    (\mathbf{F}\tran\mathbf{C}^w\mathbf{F})  \left(  \mathbf{F}\tran\mathbf{C}^m\mathbf{F} \right) ^{-\frac{1}{2}} \cdot \right. \\  
&\text{\hspace{1cm}} \left.  \left(  \mathbf{F}\tran\mathbf{C}^m\mathbf{F} \right) ^{-\frac{1}{2}}\text{Logm}_{\mathbf{F}\tran\mathbf{C}^m\mathbf{F}}    (\mathbf{F}\tran\mathbf{C}^w\mathbf{F})  \left(  \mathbf{F}\tran\mathbf{C}^t\mathbf{F} \right) ^{-\frac{1}{2}}  \right) \\
\end{aligned}
\end{equation}

After explicitly presenting the solution of $y_t^{\text{approx}}$, it seems rather sophisticated to compute $y_t^{\text{approx}}$. However, we can also notice that in the substitution of Eq.~(\ref{Eq:Lemma-0-subs}),  $\mathbf{S}_1$ and $\mathbf{C}_{\text{ref}}$ are constant once a tangent space function is found. Furthermore, if $\mathbf{C}^m$ and $\mathbf{C}^w$ can be jointly diagonalized by a properly chosen $\mathbf{F}$, the matrix multiplications in Eq.~(\ref{Eq:Lemma-0-full_substitution}) will be remarkably simplified, and an eigenvalue decomposition is no longer needed to compute $\mathbf{C}_{\text{ref}}^{-\frac{1}{2}}$, i.e., $ \left(  \mathbf{F}\tran\mathbf{C}^t\mathbf{F} \right) ^{-\frac{1}{2}} $, for test points. If the spatial filters $\mathbf{F}$ are extracted in such a manner, then $y_t^{\text{approx}}$ (Eq.~(\ref{Eq:2-13})) are simplified as:
\begin{equation}
\label{Eq:2-16}
\begin{aligned}
y_t^{\text{approx}} &= <  \text{Logm}_{\mathbf{F}\tran\mathbf{C}^m\mathbf{F}}    (\mathbf{F}\tran\mathbf{C}^w\mathbf{F}),  \\ &\text{\hspace{0.75cm}}\text{Logm}_{\mathbf{F}\tran\mathbf{C}^m\mathbf{F}}(\mathbf{F}\tran\mathbf{C}^t\mathbf{F})>\mid_{\mathbf{F}\tran\mathbf{C}^m\mathbf{F}}  \\
&= <  \text{Logm}_{\mathbf{D}^m}    (\mathbf{D}^w), \text{Logm}_{\mathbf{D}^m}(\mathbf{F}\tran\mathbf{C}^t\mathbf{F})>\mid_{\mathbf{D}^m},   \\
\end{aligned}
\end{equation}
where $\mathbf{D}^m$ is adopted to represent the filtered reference point, which is now diagonal, and $\mathbf{D}^w$ is the filtered weight matrix. 

In addition, if we can further whiten the filtered reference point, i.e., $ \mathbf{F}\tran\mathbf{C}^m\mathbf{F} = \mathbf{D}^m \Rightarrow \mathbf{I}$, then we can not only simplify $\Tr{ \mathbf{C}_{\text{ref}}^{-\frac{1}{2}} \mathbf{S}_1 \mathbf{C}_{\text{ref}}^{-\frac{1}{2}} \cdot \mathbf{C}_{\text{ref}}^{-\frac{1}{2}} \mathbf{S}_2 \mathbf{C}_{\text{ref}}^{-\frac{1}{2}} }$ to $\Tr{ \mathbf{S}_1 \mathbf{S}_2}$, but also the logarithmic map: $ \text{Logm}_{\mathbf{D}^m}  (\mathbf{D}^w) \Rightarrow \text{Logm}  (\mathbf{D}^w)$.

Therefore, by properly choosing $\mathbf{F}$, $y_t^{\text{approx}}$ can be simplified as:
\begin{equation}
\label{Eq:2-16-b}
\begin{aligned}
y_t^{\text{approx}} &= <  \text{Logm}_{\mathbf{D}^m}    (\mathbf{D}^w), \text{Logm}_{\mathbf{D}^m}(\mathbf{F}\tran\mathbf{C}^t\mathbf{F})>\mid_{\mathbf{D}^m}  \\
&= <  \text{Logm}_{\mathbf{I}}    (\mathbf{D}^w), \text{Logm}_{\mathbf{I}}(\mathbf{F}\tran\mathbf{C}^t\mathbf{F})>\mid_{\mathbf{I}}  \\
&= \Tr{  \text{Logm}_{\mathbf{I}}  ( \mathbf{D}^w) \text{Logm}_{\mathbf{I}}   ( \mathbf{F}\tran\mathbf{C}^{t}\mathbf{F}) } \\
&=  \Tr{  \text{Logm}( \mathbf{D}^w) \text{Logm} (\mathbf{F}\tran\mathbf{C}^{t}\mathbf{F}) } \\
\end{aligned}
\end{equation}

From here, we make one major assumption that the filtering matrix $\mathbf{F}$ approximately diagonalizes all $\mathbf{C}^t$. If this assumption holds , i.e., $\mathbf{F}\tran\mathbf{C}^{t}\mathbf{F} $ is a diagonally dominant matrix for all $t$, based on the Gershgorin circle theorem \cite{gershgorin1931uber} we know that 
\begin{equation}
\label{Eq:2-16-c}
\begin{aligned}
\lambda\left(  \mathbf{F}\tran\mathbf{C}^{t}\mathbf{F}  \right) \approx \text{diag}\left( \mathbf{F}\tran\mathbf{C}^{t}\mathbf{F} \right) =  \mathbf{D}^t,
\end{aligned}
\end{equation}
where $\mathbf{D}^t$ represents the diagonal matrix which only contains the diagonal elements of $ \mathbf{F}\tran\mathbf{C}^{t}\mathbf{F}$. Moreover, since $\mathbf{F}\tran\mathbf{C}^{t}\mathbf{F}$ is diagonally dominant, then following approximation can be inferred:

\begin{equation}
\label{Eq:2-16-c-ii}
\begin{aligned}
\mathbf{F}\tran\mathbf{C}^{t}\mathbf{F} \overset{}{\approx} \mathbf{D}^t
\end{aligned}
\end{equation}

Therefore, we know:
\begin{equation}
\label{Eq:2-16-d}
\begin{aligned}
\text{Logm} (\mathbf{F}\tran\mathbf{C}^{t}\mathbf{F}) \approx \text{Logm} (  \mathbf{D}^t )  
\end{aligned}
\end{equation}

After applying the approximation in Eq.~(\ref{Eq:2-16-d}) into Eq.~(\ref{Eq:2-16-b}), $y_t^{\text{approx}}$ can be simplified as:
\begin{equation}
\label{Eq:2-16-e}
\begin{aligned}
y_t^{\text{approx}}
&=  \Tr{  \text{Logm}( \mathbf{D}^w) \text{Logm} (\mathbf{F}\tran\mathbf{C}^{t}\mathbf{F}) } \\
&\approx  \Tr{  \text{Logm}( \mathbf{D}^w) \text{Logm} (\mathbf{D}^t )} \\
&=  \text{log}( \overrightarrow{d^w})^T\text{log}( \overrightarrow{d^t}), \\
\end{aligned}
\end{equation}
where $\overrightarrow{d_{(\cdot)}}$ represents the diagonal vector of $\mathbf{D}_{(\cdot)}$.

We reiterate that one primary assumption in the above simplification is that all $\mathbf{C}^t$ are roughly jointly diagonal, which is a very strong assumption. However, there is evidence for this in the fact that the projection of the physiological sources in the EEG signal to the electrodes is linear: Since the head moves very little with respect to the electrodes within a session, we can assume that the mixing (and hence unmixing) matrices stay relatively constant, even if the actual variances are non-stationary. 

The key step that enables the simplification from $y_t^{\text{approx}}  =  <\widetilde{\mathbf{S}}^w_{\perp \mathbf{F}}, \widetilde{\mathbf{S}}^t_{\perp \mathbf{F}}  >\mid_{\widetilde{\mathbf{C}}^m_{\perp \mathbf{F}}} $ to $y_t^{\text{approx}} = \text{log}( \overrightarrow{d^w})\tran\text{log}( \overrightarrow{d^t})$ is the simultaneous diagonalization  of the weight covariance matrix and the whitening of $\mathbf{C}^{m}$. The generalized eigenvalue decomposition (GED) conveniently solves both goals:
\begin{equation}
\mathbf{C}^w\mathbf{F} = \mathbf{C}^{m}\mathbf{F} \mathbf{D} \Leftrightarrow 
\left\{
\begin{array}{l}
\mathbf{F}\tran\mathbf{C}^{m}\mathbf{F} = \mathbf{I}\\
\mathbf{F}\tran\mathbf{C}^w\mathbf{F} = \mathbf{D} 
\end{array}
\right. , 
\label{Eq:2-17}
\end{equation}
where the $\mathbf{F}$ is named as tangent space spatial filter (TSSF) and $\mathbf{D}$ is the corresponding eigenvalues. Importantly, to ensure that Eq.~(\ref{Eq:2-17}) will hold, the order of $\mathbf{C}^{m}$ and $\mathbf{C}^{w}$ in the GED equation cannot be switched. 

Now, since $y_t^{\text{approx}}$ can be drastically simplified as long as $\mathbf{F}$ are extracted with the GED manner, when looking back to the objective function for extracting optimal filters, i.e., $\mathbf{F}_K^{\ast} = \underset{ \mathbf{F}_K \in \Re^{N \times K}}{\arg \max} \sum^T(\hat{y}_t- y_t^{\text{approx}}\mid_{\mathbf{F}_K} )^2 $, the last remaining obstacle is the true predicted label $\hat{y}_t$. We next prove that the equivalence between $\hat{y}_t$ and $y_t^{\text{approx}}$ will hold under some conditions, as seen in \textit{theorem 1}.

\textbf{\textit{Theorem 1: Equivalence between true and approximated decision function}}
\begin{equation}
\label{Eq:theorem-1}
\begin{aligned}
\hat{y}_t \equiv y_t^{\text{approx}}, &\text{ iff $\mathbf{F}$ is extacted via GED($ \mathbf{C}^w$, $ \mathbf{C}^m$)} \\
&\text{ and full rank.}
\end{aligned}
\end{equation}

\textbf{\textit{Proof:}}
For convenience, we first list some properties of linear algebra in the tangent space that will be convenient in the proof. All properties are proved in Appendix \ref{Appendix A}.
\begin{enumerate} \label{List_prev}
    \item $\mathbf{F}, \mathbf{D}    = \text{GED}(\mathbf{C}^w, \mathbf{C}^m) = \text{ED}(\mathbf{C}^{-m} \mathbf{C}^w)$, where $\mathbf{F}\tran \mathbf{F} \neq \mathbf{I}$
    \item $\mathbf{V}, \mathbf{D}  = \text{ED}(   \mathbf{C}^{-\frac{m}{2}} \mathbf{C}^w  \mathbf{C}^{-\frac{m}{2}} )  $, where $ \mathbf{V}\tran \mathbf{V} = \mathbf{I}$
    \item  $\mathbf{V} = \mathbf{C}^{\frac{m}{2}} \mathbf{F}  \Rightarrow \mathbf{F}\tran \mathbf{C}^{m}\mathbf{F} = \mathbf{I}$ 
\end{enumerate}
Furthermore, we would like to quote one significant lemma in which the following equivalence holds based on the property of affine-invariance Riemannian metric. The corresponding proof is shown in Appendix {\ref{GED_approx_lemma2}}.

\textbf{\textit{Lemma 2: Equivalence of logarithm mapping}}
\begin{equation}
\begin{aligned}
\label{Eq:appendix-B-Lemma-1}
\text{Logm}(\mathbf{V}\tran\mathbf{A}\mathbf{V}) = \mathbf{V}\tran\text{Logm}(\mathbf{A})\mathbf{V} \\ \text{, iff }  \mathbf{V}\tran \mathbf{V} =  \mathbf{I} \text{ and $\mathbf{A}$ is SPD.}  
\end{aligned}
\end{equation}

Since spatial filter $ \mathbf{F}$ and the eigenvalue matrix $ \mathbf{D}$ are extracted via the GED($ \mathbf{C}^w$, $ \mathbf{C}^m$), we have:
\begin{equation}
\label{Eq:appendix-B-2}
\begin{aligned}
&\widetilde{\mathbf{C}}^m_{\perp \mathbf{F}}  = \mathbf{F}\tran\mathbf{C}^m\mathbf{F} = \mathbf{I}\\ 
&\widetilde{\mathbf{C}}^w_{\perp \mathbf{F}}  = \mathbf{F}\tran\mathbf{C}^w\mathbf{F} = \mathbf{D}\\ 
\end{aligned}
\end{equation}

On the one hand, Eq.~(\ref{Eq:2-16}) and (\ref{Eq:2-16-b}) show that the approximated decision function can be reformulated as:
\begin{equation}
\label{Eq:appendix-B-3}
\begin{aligned}
y_t^{\text{approx}} = \Tr{  \text{Logm}   (  \mathbf{D})  \text{Logm} ( \mathbf{F}\tran\mathbf{C}^{t}\mathbf{F} )     } 
\end{aligned}
\end{equation}

On the other hand, based on the previous proved and quoted lemmas as well as the definitions, the true decision function can be expressed as: 
\begin{equation}
\label{Eq:appendix-B-4}
\begin{aligned}
\hat{y}_t &= <\mathbf{S}^w, \mathbf{S}^t>\mid_{\mathbf{C}^m} \\
&\overset{\text{Lemma } 1}{=} <\mathbf{C}^{-\frac{m}{2}}\mathbf{S}^w\mathbf{C}^{-\frac{m}{2}}, \mathbf{C}^{-\frac{m}{2}}\mathbf{S}^t\mathbf{C}^{-\frac{m}{2}}>\mid_{\mathbf{I}} \\
&= \Tr{\mathbf{C}^{-\frac{m}{2}}\mathbf{S}^w\mathbf{C}^{-\frac{m}{2}} \cdot \mathbf{C}^{-\frac{m}{2}}\mathbf{S}^t\mathbf{C}^{-\frac{m}{2}}}\\
&\overset{\text{Eq.~(\ref{Eq:2-8})}}{=}    \text{Tr}\left( \mathbf{C}^{-\frac{m}{2}}\left(\text{Logm}_{\mathbf{C}^m}(\mathbf{C}^w) \right) \mathbf{C}^{-\frac{m}{2}} \cdot  \right. \\  
& \text{\hspace{0.8cm} }\left.\mathbf{C}^{-\frac{m}{2}}\left(\text{Logm}_{\mathbf{C}^m}(\mathbf{C}^t) \right) \mathbf{C}^{-\frac{m}{2}}  \right)  \\
&\overset{\text{Eq.~(\ref{Eq:2-3})}}{=} \text{Tr}\left( \mathbf{C}^{-\frac{m}{2}}\left(\mathbf{C}^{+\frac{m}{2}}  \text{Logm}\left( \mathbf{C}^{-\frac{m}{2}} \mathbf{C}^{w} \mathbf{C}^{-\frac{m}{2}}\right)    \mathbf{C}^{+\frac{m}{2}}    \right) \mathbf{C}^{-\frac{m}{2}} \cdot  \right. \\  
& \text{\hspace{0.8cm} }\left.\mathbf{C}^{-\frac{m}{2}}\left(\mathbf{C}^{+\frac{m}{2}}  \text{Logm}\left( \mathbf{C}^{-\frac{m}{2}} \mathbf{C}^{t} \mathbf{C}^{-\frac{m}{2}}\right)    \mathbf{C}^{+\frac{m}{2}}    \right) \mathbf{C}^{-\frac{m}{2}} \right)  \\
&= \text{Tr}\left(\text{Logm}\left( \mathbf{C}^{-\frac{m}{2}} \mathbf{C}^{w} \mathbf{C}^{-\frac{m}{2}}\right)  \cdot \text{Logm}\left( \mathbf{C}^{-\frac{m}{2}} \mathbf{C}^{t} \mathbf{C}^{-\frac{m}{2}}\right)  \right)  \\
\end{aligned}
\end{equation}

As indicated by point 2) of the list at the beginning of this proof (refer as List 2. in the following context), we know that $\mathbf{V}, \mathbf{D}  = \text{ED}(   \mathbf{C}^{-\frac{m}{2}} \mathbf{C}^w  \mathbf{C}^{-\frac{m}{2}} )  $, where $ \mathbf{V}\tran \mathbf{V} = \mathbf{I}$. Therefore, the results from Eq.~(\ref{Eq:appendix-B-4}) can be further derived as:
\begin{equation}
\label{Eq:appendix-B-5}
\begin{aligned}
\hat{y}_t 
&= \text{Tr}\left(\text{Logm}\left( \mathbf{C}^{-\frac{m}{2}} \mathbf{C}^{w} \mathbf{C}^{-\frac{m}{2}}\right)  \cdot \text{Logm}\left( \mathbf{C}^{-\frac{m}{2}} \mathbf{C}^{t} \mathbf{C}^{-\frac{m}{2}}\right)  \right)  \\
&\overset{\text{List } 2).}{=}  \text{Tr}\left(\text{Logm}\left( \mathbf{V}\mathbf{D}\mathbf{V}\tran\right)  \cdot \text{Logm}\left( \mathbf{C}^{-\frac{m}{2}} \mathbf{C}^{t} \mathbf{C}^{-\frac{m}{2}}\right)  \right)  \\
&\overset{\text{Lemma } 2}{=}  \text{Tr}\left(  \mathbf{V} \text{Logm}\left(\mathbf{D}\right) \mathbf{V}\tran \cdot \text{Logm}\left( \mathbf{C}^{-\frac{m}{2}} \mathbf{C}^{t} \mathbf{C}^{-\frac{m}{2}}\right)  \right)  \\
&\underset{\text{invariant}}{\overset{\text{Trace cyclic}}{=}} \text{Tr}\left( \text{Logm}\left(\mathbf{D}\right)\cdot \mathbf{V}\tran  \text{Logm}\left( \mathbf{C}^{-\frac{m}{2}} \mathbf{C}^{t} \mathbf{C}^{-\frac{m}{2}}\right)   \mathbf{V} \right)  \\
&\overset{\text{Lemma } 2}{=} \text{Tr}\left( \text{Logm}\left(\mathbf{D}\right)\cdot \text{Logm}\left( \mathbf{V}\tran  \mathbf{C}^{-\frac{m}{2}} \mathbf{C}^{t} \mathbf{C}^{-\frac{m}{2}}  \mathbf{V} \right)   \right)  \\
&\overset{\text{List } 3).}{=} \text{Tr}\left( \text{Logm}\left(\mathbf{D}\right)\cdot \right. \\ 
&\text{\hspace{1.4cm}} \left. \text{Logm}\left( \left( \mathbf{C}^{+\frac{m}{2}} \mathbf{F} \right) \tran  \mathbf{C}^{-\frac{m}{2}} \mathbf{C}^{t} \mathbf{C}^{-\frac{m}{2}} \left( \mathbf{C}^{+\frac{m}{2}} \mathbf{F} \right)\right)   \right)  \\
&= \text{Tr}\left( \text{Logm}\left(\mathbf{D}\right)\cdot \text{Logm}\left( \mathbf{F} \tran  \mathbf{C}^{+\frac{m}{2}}\mathbf{C}^{-\frac{m}{2}} \mathbf{C}^{t} \mathbf{C}^{-\frac{m}{2}} \mathbf{C}^{+\frac{m}{2}} \mathbf{F} \right)   \right)  \\
&= \text{Tr}\left( \text{Logm}\left(\mathbf{D}\right)\cdot \text{Logm}\left( \mathbf{F} \tran  \mathbf{C}^{t} \mathbf{F} \right)   \right)  \\
&= y_t^{\text{approx}}
\end{aligned}
\end{equation}

Therefore, as a summary, $\hat{y}_t \equiv y_t^{\text{approx}}$, if and only if when $\mathbf{F}$ is extracted via GED($ \mathbf{C}^w$, $ \mathbf{C}^m$) and $\mathbf{F}$ is with full rank.

\textbf{Q.E.D}

By leveraging this equivalence, the objective function in Eq.~(\ref{Eq:2-14}) can be reformulated as: 

\begin{equation}
\label{Eq:2-15}
\begin{aligned}
\mathbf{F}_K^{\ast} = \underset{ \mathbf{F}_K   \in \Re^{N \times K } }{\arg \min} \sum^T(y_t^{\text{approx}}\mid_{\mathbf{F}}- y_t^{\text{approx}}\mid_{\mathbf{F}_K} )^2 
\end{aligned}
\end{equation}

Since the $\mathbf{F}_K$ is known as the subset of $\mathbf{F}$ which is extracted from the $\text{GED}(\mathbf{C}^w,\mathbf{C}^m)$, the optimization problem in Eq.~(\ref{Eq:2-15}) is then equivalent to the problem of ordering the columns of $\mathbf{F}$.

This problem can be tackled by observing the result in Eq.~(\ref{Eq:2-16-e}), which states that as long as the filtered input data $\widetilde{\mathbf{C}}^{t}_{\perp \mathbf{F}}$ is roughly diagonal, the linear functions in the Riemannian tangent space can be approximated by linear functions of the log-variances of the filtered data. More importantly, the coefficients of this approximated linear function are simply the log-eigenvalues after the GED is solved. i.e., $\text{log}( \overrightarrow{d})$.  Thus, standard techniques for determining the most important variables in a linear regression problem can be used. For simplicity's sake, we use the absolute values of the regression coefficients as markers of their importance to the function.

\paragraph{Intuitive Explanation}
One very common and effective technique across domains is whitening data. By decorrelating the different channels, constructed features are often more distinct and predictive. However, whitening has a fundamental flaw, in that there are arbitrarily many whitening matrices that are possible since the covariance of whitened data is invariant to rotations. One explanation for the finding above is that the GED can be decomposed into a whitening transform and a subsequent rotation. The whitening is with respect to the data, and the rotation is chosen based on the weight matrix. Therefore this technique can be considered a particular choice of data whitening that simultaneously preserves the information of a function in the tangent space. 

\subsection{The Classification based on the TSSF} \label{sec_TSSF_clf}
As a feature extraction method, spatial filtering always requires a classifier to deal with the processed features, which often requires an extra optimization step. One method to use the proposed TSSF is like any other feature reduction technique, fitting a classifier after the spatial filtering step. However, another advantage brought by the linear approximation function of TSSF is that this secondary training can be skipped, which means it is possible to further reduce the computational time for TSSF.

From Eq.~(\ref{Eq:2-16-e}), we notice that this function is actually a linear regressor using the log-eigenvalues of the GED as the regression weights. Therefore, we can directly input the filtered data into this regressor to obtain the predicted value. This method is named as one-step classification in our paper, and the ordinary way to classify the data is named as two-steps classification, i.e., filtering and classifying.

\subsection*{Example 1: Tangent Space Spatial Filter - The Generalization of CSP} \label{example_CSP}
As a data-driven spatial filter, it is inevitable to compare the performance of TSSF and CSP, especially considering the great impact of the latter in the BCI field. Instead of merely comparing the performance of both filters, we also prove that CSP is a special case of TSSF. To begin with the proof of their relationship, let us first review the TSSF.

The solutions of TSSF are obtained via solving the $\text{GED}(\mathbf{C}^w,\mathbf{C}^m)$ as described in Section \ref{sec_TSSF}. Moreover, the equivalent solution of eigenvectors can also be extracted by solving $\text{GED}(\mathbf{S}^w,\mathbf{C}^m)$, i.e.,: 

\begin{equation}
\label{Eq:2-17-a}
\begin{aligned}
\mathbf{F}, \mathbf{D}  &= \text{GED}(\mathbf{C}^w, \mathbf{C}^m) \\
\mathbf{F},  \text{Logm}\left( \mathbf{D}\right)   &= \text{GED}(\mathbf{S}^w, \mathbf{C}^m),\\
\end{aligned}
\end{equation}
the proof of which can be referred in Section \ref{Appendix A}.

Furthermore, when the classifier on the tangent space is specified as the Fisher LDA classifier \cite{bishop2006pattern}, the weight vector on the tangent space is as expressed in Eq.(\ref{Eq:2-18}). The corresponding proof can be found in \cite{bishop2006pattern}.
\begin{equation}
\label{Eq:2-18}
\begin{aligned}
\overrightarrow{w}_{\text{LDA}}  &= \mathbf{S}_{within}^{-1}( \mathbf{\mu}^{(+)}- \mathbf{\mu}^{(-)}) \\
\mathbf{S}_{within}  &= \sum_{a\in\{+, -\} }^{} \sum_{t\in ^{(a)}}^{} \left( \overrightarrow{s^t}  - \mathbf{\mu}^{(a)}\right) \left( \overrightarrow{s^t} - \mathbf{\mu}^{(a)}\right)\tran,
\end{aligned}
\end{equation}
where $\mu = \frac{1}{T}\sum_{t=1}^{T} \overrightarrow{s^t} $ and $\mathbf{\mu}^{(a)}, a \in \{+,-\}$ are the within class mean for the tangent vectors and $ \mathbf{S}_{within}$ is the within scatter matrix.

Under the special case that the $\mathbf{S}_{within}$ is equal to the identity matrix $\mathbf{I}$, the weight vector of LDA classifier is simplified as:
\begin{equation}
\label{Eq:2-20}
\begin{aligned}
\overrightarrow{w}_{\text{LDA}}  =  \mathbf{\mu}^{(+)}- \mathbf{\mu}^{(-)} \in \Re^{\frac{C\left(C+1 \right) }{2} \times 1} \\
\end{aligned}
\end{equation}

Based on the reverse operation of $\vt{\cdot}$ (Eq.~(\ref{Eq:2-4})), the equivalent formulation of Eq.~(\ref{Eq:2-20}) in matrix format is:
\begin{equation}
\label{Eq:2-21}
\begin{aligned}
\mathbf{S}^{w_{\text{LDA}} } =  \overline{\mathbf{S}^{(+)}} -  \overline{\mathbf{S}^{(-)}}, \\
\end{aligned}
\end{equation}
where $\overline{\mathbf{S}^{(\cdot)}}$ is the arithmetic within-class mean for project points on the tangent space. Moreover, assuming the special situation holds in which the between-class Euclidean mean difference of the covariances is the exponential transform of the between-class Euclidean mean difference of tangent space points, i.e., 
\begin{equation}
\label{Eq:2-22}
\begin{aligned}
\overline{\mathbf{C}^{(+)}} -  \overline{\mathbf{C}^{(-)}} = \text{Expm}_{\mathbf{C}^m}\left( \overline{\mathbf{S}^{(+)}} -  \overline{\mathbf{S}^{(-)}}\right) 
\end{aligned}
\end{equation}

combining the special LDA classifier with the conclusion drawn from equation(\ref{Eq:2-17-a}), the solution of TSSF can be further formulated as:
\begin{equation}
\label{Eq:2-23}
\begin{aligned}
\text{GED}(\mathbf{C}^w,\mathbf{C}^m) & \overset{\text{LDA as clf.}}{\Rightarrow} \text{GED}(\mathbf{C}^{w_{\text{LDA}} }, \mathbf{C}^m) \\
&\overset{\text{Eq. (\ref{Eq:2-17-a})}}{=} \text{GED}(\mathbf{S}^{w_{\text{LDA}} }, \mathbf{C}^m) \\
&\overset{\text{Eq. (\ref{Eq:2-21})}}{=} \text{GED}(\overline{\mathbf{S}^{(+)}} -  \overline{\mathbf{S}^{(-)}}, \mathbf{C}^m) \\
&\overset{\text{Eq. (\ref{Eq:2-17-a}\& \ref{Eq:2-22})}}{=}  \text{GED}(\overline{\mathbf{C}^{(+)}} -  \overline{\mathbf{C}^{(-)}}, \mathbf{C}^m) , \\
\end{aligned}
\end{equation}
where $\text{GED} (\mathbf{A}, \mathbf{B})$ in above equations represents the corresponding eigenvectors, i.e., $\mathbf{V} = \text{GED} (\mathbf{A}, \mathbf{B})$ and $\mathbf{A}\mathbf{V} = \mathbf{B}\mathbf{V}\mathbf{\Gamma}$ ($\mathbf{\Gamma}$ is the matrix of generalized eigenvalues).

In addition, if we further replace the Fr\'{e}chet mean $\mathbf{C}^m$ in Eq.~(\ref{Eq:2-23}) with arithmetic mean $\overline{\mathbf{C}^m}$, we will have:
\begin{equation}
\label{Eq:2-25}
\begin{aligned}
\text{GED}(\mathbf{C}^w,\mathbf{C}^m) & \overset{\text{Eq. (\ref{Eq:2-23})}}{\Rightarrow}  \text{GED}( \overline{\mathbf{C}^{(+)}} -  \overline{\mathbf{C}^{(-)}},  \mathbf{C}^m) \\
& \overset{}{\Rightarrow}  \text{GED}( \overline{\mathbf{C}^{(+)}} -  \overline{\mathbf{C}^{(-)}},  \overline{\mathbf{C}^{m}} ) \\
& \overset{}{\Rightarrow}  \text{GED}( \overline{\mathbf{C}^{(+)}} -  \overline{\mathbf{C}^{(-)}},  \frac{\overline{\mathbf{C}^{(+)}} +  \overline{\mathbf{C}^{(-)}}}{2}) \\
& \underset{\text{invariance}}{ \overset{\text{Scaling }}{\Rightarrow}}   \text{GED}( \overline{\mathbf{C}^{(+)}} -  \overline{\mathbf{C}^{(-)}},\overline{\mathbf{C}^{(+)}} +  \overline{\mathbf{C}^{(-)}}) \\
\end{aligned}
\end{equation}

By combining the definition of CSP from the discriminative perspective as described in Eq.~(\ref{Eq:2-26}) and the equivalence as shown in Eq.~(\ref{Eq:2-25}), we are able to conclude the relationship between CSP and TSSF as:

\begin{equation}
\label{Eq:2-28}
\begin{aligned}
\mathbf{F}_{\text{TSSF}} & \overset{\text{Def.}}{\Rightarrow} \text{GED}(\mathbf{C}^w,\mathbf{C}^m) \\
&  \overset{\text{Eq. (\ref{Eq:2-25})}}{\Rightarrow}  \text{GED}( \overline{\mathbf{C}^{(+)}} -  \overline{\mathbf{C}^{(-)}},\overline{\mathbf{C}^{(+)}} +  \overline{\mathbf{C}^{(-)}} ) \\
& \overset{\text{Eq. (\ref{Eq:2-26})}}{\equiv}     \text{GED}(\mathbf{C}_d,\mathbf{C}_c)  \\
& \overset{\text{Eq. (\ref{Eq:2-27})}}{\Rightarrow}  \mathbf{F}_{\text{CSP}}
\end{aligned}
\end{equation}

Namely, CSP is the representation of TSSF when LDA is chosen as the classifier on the tangent space, and the within-class scatter matrix is assumed to be the identity. One important caveat is the exponential relationship of class mean subtraction, as shown in the Eq.~(\ref{Eq:2-23}), which is not necessarily true. %

One related work we would like to mention in this example is \cite{Barachant2010_CSP_revisit}, in which Barachant \textit{et al} replaced the arithmetic mean with the Fr\'{e}chet mean in CSP. One crucial component of our equivalence in Section \ref{example_CSP} is the relationship between $\overline{\mathbf{C}^{(\cdot)}}$ and $\overline{\mathbf{S}^{(\cdot)}}$. We assume that they are related by the exponential transform, but that is likely not true if $\overline{\mathbf{C}^{(\cdot)}}$ is computed as the arithmetic mean of the covariance matrices in a given class, due to the swelling effect \cite{arsigny2006log}. Since the Fr\'{e}chet mean is a much better proxy of common activities across trials, the proposed Riemannian CSP is a far better approximation of LDA in the tangent space, and Barachant \textit{et al} also show increased performance and robustness with this alteration. 

\subsection{Summary of the extraction and application of TSSF}
\label{sec:summary}
For practitioners interested in using the proposed TSSF framework, we summarize its procedures in this subsection. Generally, the usage of TSSF can be divided into two stages: how to extract spatial filters and how to use the spatially filtered signals for BCIs. Therefore, the corresponding algorithms are introduced below and summarized into pseudocode separately. To link each algorithm's description with its pseudocode, we adopt the abbreviation that A1-1 denotes the Step-1 of Algorithm 1.

\subsubsection{Extraction of TSSF}
To extract the TSSF, the input data should be bandpass filtered data already epoched into trials, and the choice of the linear model on the tangent space is supposed to be defined beforehand. Subsequently, the covariance matrices are estimated based on the input trialwise EEG signal and their Fr\'{e}chet mean is computed to use as the reference point for the tangent space projection (A1-1 and A1-2). After finding the Fr\'{e}chet mean, all covariance matrices are projected onto the tangent space and vectorized into tangent vectors (A1-3 and A1-4). Afterward, by using these tangent vectors the linear model is trained, the weights are obtained (A1-5) and reshaped into a symmetric matrix, and the equivalent weight covariance matrix on the manifold is computed via the exponential transform (A1-6). 
Next, the full-rank filter matrix of TSSF, as well as the regression coefficients for one-step classification, are obtained by solving the GED problem (A1-7) and sorting based on the absolute value of the logarithm of the eigenvalues in descending order (A1-8 and A1-9). At last, based on the predefined parameter that how many filter components are needed, the first $K$ components of the full and sorted filter matrix are extracted, and the same is done with the regression coefficients (A1-10).

\begin{algorithm}[h!]
    \SetAlgoLined
    \KwData{Bandpass filtered trialwise data $\mathbf{X} \in \Re^{C \times N \times T}$, loss function for linear model \textit{L}}
    \KwResult{TSSF and regression coefficients with K components: \\ \hskip 1.5cm $\mathbf{F}_K \in \Re^{C \times K}, \overrightarrow{\beta}_K \in \Re^{k \times 1}$ }
    \Begin{
        1. Compute the covariance matrices:\\ $\mathbf{C}^t = \mathbf{X}^t (\mathbf{X}^t) \tran, \forall t \in [1, \cdots, T]$. 
        
        2. Compute the Fr\'{e}chet mean: \\$\mathbf{C}^m =\underset{\mathbf{A} \in \mathbf{C}  }{\arg \min} \sum_{t=1}^{T} d_{\text{AIRM}}^2\left( \mathbf{A},  \mathbf{C}^t \right)$
        
        3. Project onto tangent space:\\ $\mathbf{S}^t= \text{Logm}_{\mathbf{C}^m }\left( \mathbf{C}^t\right)$, $\forall t \in [1, \cdots, T]$
        
        4. Compute tangent vectors:\\ $\overrightarrow{s^t} = \vt{\mathbf{S}^t}, \forall t \in [1, \cdots, T]$ and $\mathbf{s} = [\overrightarrow{s^1}, \cdots, \overrightarrow{s^T}] \in \Re^{\frac{C(C+1)}{2}\times T}$
        
        5. Fit linear model:\\ $\overrightarrow{w} = \text{argmin}_{\overrightarrow{w}}L(\mathbf{s}, \overrightarrow{ w })\in \Re^{\frac{C(C+1)}{2} \times 1}$
        
        6. Project weights onto manifold:\\ $\overrightarrow{w} \overset{\text{unvec}(\cdot) }{\Rightarrow}\mathbf{S}^{w} \overset{\text{Expm}_{\mathbf{C}^m }(\cdot)}{\Rightarrow} \mathbf{C}^{w} \in \Re^{C \times C}$
        
        7. Solve the Generalized Eigenvalue Decomposition (GED) problem:\\ $\overrightarrow{d},\mathbf{V} = \text{GED} \left(\mathbf{C}^{w}, \mathbf{C}^{m}\right) $
        
        8. Get the sorted index based on the value of $\overrightarrow{d}$:\\ $\text{inds} = \text{sort}\left( \left| \text{log}(\overrightarrow{d} ) \right| \right) $
        
        9. Obtain the sorted TSSF and regression coefficients: \\ \hskip 1.5cm $\mathbf{F} = \mathbf{V}[:, \text{inds}] \in \Re^{C \times C}, \overrightarrow{\beta} =  \text{log}(\overrightarrow{d}[\text{inds}] )  \in \Re^{C \times 1} $
        
        10. Extract the first $K$ components:\\ $\mathbf{F}_K  = \mathbf{F}[:, :K], \overrightarrow{\beta}_K = \overrightarrow{\beta}[:K]$
    }
    \caption{Extraction of Tangent Space Spatial Filter (TSSF)}
    \label{Algo-1}
\end{algorithm}

\subsubsection{Application of TSSF}
Once the TSSF are extracted, there are some options regarding how to generate features and use the trained linear model. The first step is to apply the extracted spatial filters onto the trialwise data (A2-1). Subsequently, there are three types of features which can be generated from the filtered data: the log-variance of filtered data (A2-2.a)), the diagonal vector of the logarithm of filtered covariance matrices (A2-2.b)) and the full tangent vector computed based on filtered covariance matrices (A2-2.c)). These three types of features and their descriptions, as well as corresponding abbreviations, are summarized in Table \ref{Feat_table}. 

\begin{table}[htb!]
    \centering
    \begin{tabular}{|c|c|c|}
        \hline
        \textbf{Formulation} & \textbf{Description} &  \textbf{Abbreviation}
        \\ \hline
        $ \text{log} \left( \text{diag}(\widetilde{\mathbf{C}}^t_{\perp \mathbf{F}})\right) $                                                                & Log-variance                         &Log-var                                                                    \\ \hline
        \multirow{2}{*}{$\text{diag} \left( \text{Logm}_{\mathbf{I}}(\widetilde{\mathbf{C}}^t_{\perp \mathbf{F}})\right) $}                                  & \multirow{2}{*}{\begin{tabular}[c]{@{}c@{}}Diagonal of logarithm of \\ covariance matrices\end{tabular}} & \multirow{2}{*}{Diag. log-cov} \\
        &                                                                                                                                                      &                                                                                                          \\ \hline
        \multirow{2}{*}{$\text{vec}\left(\text{Logm}_{\widetilde{\mathbf{C}}^{m}_{\perp \mathbf{F}}}(\widetilde{\mathbf{C}}^t_{\perp \mathbf{F}}) \right) $} & \multirow{2}{*}{\begin{tabular}[c]{@{}c@{}}Logarithm of covariance\\ matrices\end{tabular}}      &    \multirow{2}{*}{Log-cov}        \\
        &                                                                                                                                                      &                                                                                                          \\ \hline
    \end{tabular}
    \caption{Summary of classifiable features}
    \label{Feat_table}
\end{table}

After obtaining the classifiable features, the last step is to classify them. As described in Section \ref{sec_TSSF_clf}, two possible classification algorithms can be applied: one-step classification and two-steps classification. One thing that should be noted is that one-step classification can only be applied to the diagonal elements based features, namely features from (A2-2.b)) and (A2-2.c)). For the one-step classification, the inner product between regression coefficients and features are computed, and the label is taken as the sign of the result in binary classification problems. For the two-steps classification, a second classifier is chosen and fitted with the features from the training set (A2-2.b).i)). After that, test data to be classified can be classified by this trained second classifier.
\begin{algorithm}[h!]
    \SetAlgoLined
    \KwData{Test trialwise data $\mathbf{X} \in \Re^{C \times N \times T}$, Second classifier $\textit{Clf}_2$ if needed}
    \KwResult{TSSF and regression coefficients: $\mathbf{F}_K \in \Re^{C \times K}, \overrightarrow{\beta}_K \in \Re^{K \times 1}$ }
    \Begin{
        1. Filter the test data: $\widetilde{\mathbf{X}}_{\perp \mathbf{F}_K} = \mathbf{F}_K\tran \mathbf{X} \in \Re^{K \times N} $
        
        2. Compute features (several options are provided, only choose one):
        
        \hskip 0.5cm a). $ \overrightarrow{e} = \text{log} \left( \text{var} \left( \widetilde{\mathbf{X}}_{\perp \mathbf{F}_K} \right) \right) \in \Re^{K \times 1}$
        
        \hskip 0.5cm b). $\widetilde{\mathbf{X}}_{\perp \mathbf{F}_K} \overset{\text{Cov}(\cdot) }{\Rightarrow} \widetilde{\mathbf{C}}_{\perp \mathbf{F}_K}  \overset{ }{\Rightarrow}  \overrightarrow{e} = \text{diag}\left( \text{Logm}\left( \widetilde{\mathbf{C}}_{\perp \mathbf{F}_K}  \right) \right)  \in \Re^{K \times 1}$
        
        \hskip 0.5cm c). $\widetilde{\mathbf{X}}_{\perp \mathbf{F}_K} \overset{\text{Cov}(\cdot) }{\Rightarrow} \widetilde{\mathbf{C}}_{\perp \mathbf{F}_K}  \overset{ }{\Rightarrow} \overrightarrow{e} = \text{vec}\left( \text{Logm}\left( \widetilde{\mathbf{C}}_{\perp \mathbf{F}_K}  \right) \right)   \in \Re^{\frac{K(K+1)}{2} \times 1}$
        
        3. Return label (several options are provided, only choose one):
        
        \hskip 0.5cm a). One-step classificaiton (only applicaple for features from 2.a) or 2.b)): 
        
        \hskip 1.0cm  i).  $\hat{y} =  \text{sgn}(\overrightarrow{\beta}_K \tran \overrightarrow{e})$ 
        
        \hskip 0.5cm b). Two-steps classificaiton (applicaple for all features): 
        
        \hskip 1.0cm i).  Use a set of $\overrightarrow{e}$ from training datasets to fit a second classifier $\textit{Clf}_2$
        
        \hskip 1.0cm ii). Use the fitted classifier to classify the testing datasets and obtain the predicted label.
        
        \KwRet predicted label
        
    }
    \caption{Feature generation and classification}
    \label{Algo-2}
\end{algorithm}

\subsection{Experimental Setup}
Now that we have shown the theoretical validity of tangent space spatial filtering, we move on to our empirical results. We base our experimental setup on a recently released open-source benchmark, which is known as \textit{Mother of all BCI Benchmark (MOABB)} \cite{jayaram2018moabb}. After that, we first fix the experimental paradigm as left-hand versus right-hand motor imagery because the corresponding neurophysiological knowledge, as well as the activated neuronal sources, are well studied. Furthermore, the analysis is restricted to the $\alpha$- and $\beta$-bands (8Hz $\sim$ 32Hz) based on neurophysiological knowledge. Also, all channels are utilized except for the electrooculography (EOG) channel. 
Based the chosen paradigm, we tried to adopt all eight available datasets in the MOABB, as summarized in Table \ref{T2}; however, as indicated in the Table \ref{T2}, the dataset \textit{BNCI 2014-004} is excluded from the analysis (marked in red in Table \ref{T2}) due to having only three electrodes.

\begin{table*}[htb!]
    \caption{\protect \centering  Overview of all adopted datasets with left-hand versus right-hand motor imagery paradigm. The dataset marked in red color has only 3 channels and is hence excluded from this analysis.}%
    \label{T2}
    \centering
    \begin{tabular}{|c|c|c|c|c|c|}
        \hline
        {Dataset Name} & {\#Channels} &{\#Subjects} & {\#Sessions} \\ 
        \hline
        BNCI 2014-001                        & 22                          & 9                           & 2                           \\
        \textcolor{red}{BNCI 2014-004 }              &\textcolor{red}{3}                           & \textcolor{red}{9}       & \textcolor{red}{5}                   \\
        Cho et al. 2017              & 64                          & 49                          & 1                           \\
        Munich Motor Imagery   & 128                         & 10                          & 1                           \\
        Physionet Motor Imagery         & 64                          & 109                         & 1                           \\
        Shin et al. 2017              & 25                          & 29                          & 3                           \\
        Weibo et al. 2014               & 60                          & 10                          & 1                           \\
        Zhou et al. 2016             & 14                          & 4                           & 3                           \\ \hline
    \end{tabular}
\end{table*}

After the bandpass filtering, the covariance matrices are first estimated from the trial-wise data via the empirical covariance estimator. Subsequently, three algorithms are employed to generate feature: CSP, TSSF, and standard Riemannian tangent space methods. TSSF based features are further subdivided into three types depending on the degree of approximation as summarized in Table \ref{Feat_table}, and two methods, namely one-step and two-steps classification as described in Section \ref{sec_TSSF_clf}. The difference between them is the choice of the second classifier: either fitting a new classifier after spatial filter generation (two-steps) or employing the eigenvalues from the GED solution as linear regression coefficients. These classification methods are summed up in Table \ref{Clf_table}. For CSP and standard Riemannian features, the L2-regularized SVM classifier is used as a classifier.

\begin{table}[htb!]
    \centering
    \begin{tabular}{|c|c|c|}
        \hline
        \textbf{Name}             & \textbf{First classifier}        & \textbf{Second Classifier}                                                                                                                           \\ \hline
        \multirow{3}{*}{One-step} & \multirow{3}{*}{L2 Regularized SVM} & \multirow{3}{*}{\begin{tabular}[c]{@{}c@{}}Linear regression based on \\ the eigenvalue of GED\\  (refer to section \ref{sec_TSSF_clf}) \end{tabular}} \\
        &                                  &                                                                                                                                                      \\
        &                                  &                                                                                                                                                      \\ \hline
        Two-steps                  & L2 Regularized SVM                  & L2 Regularized SVM                                                                                                                                      \\ \hline
    \end{tabular}
    \caption{Summary of classifiers. For all regularized SVM listed above, the parameters are found by grid search \cite{pedregosa2011scikit}.}
    \label{Clf_table}
\end{table}

The motivation of selecting a regularized SVM as the first classifier to generate weight vectors on the tangent space is inspired by the results from \cite{jayaram2018moabb}, in which the combination of regularized SVM and Riemannian methods has been validated as the best among all benchmarked pipelines. For choosing hyperparameters, a grid search \cite{pedregosa2011scikit} is employed to find the optimal value within the range from 0.01 to 100.

\begin{figure}[htb!]
    \centering
    \includegraphics[width=.96\linewidth]{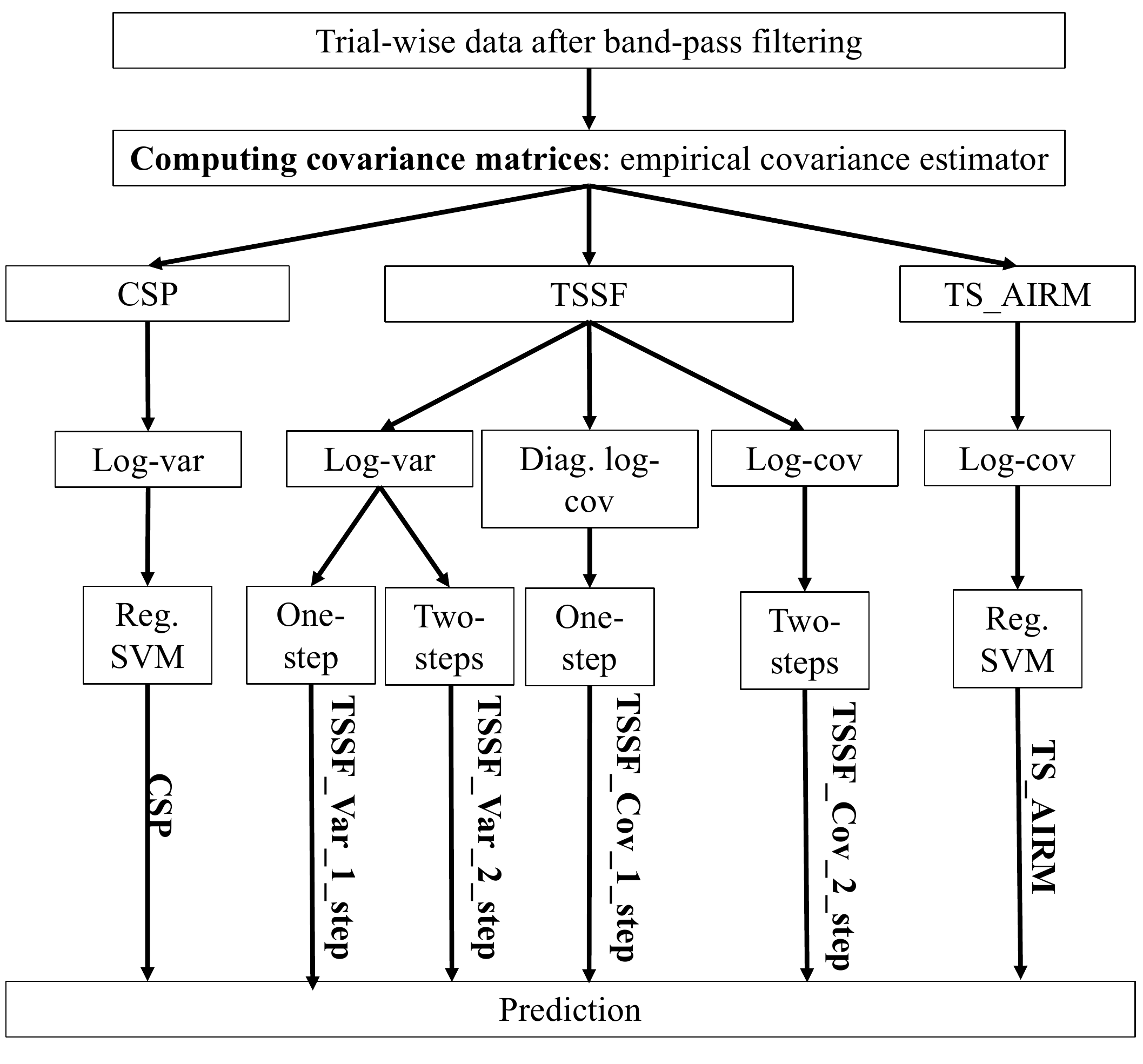}
    \caption{All tested pipelines in this paper. The annotated text above the line linked between classifiers and predictions is the abbreviation of the corresponding pipeline and Reg. SVM is the abbreviation of L2 regularized SVM.}
    \label{Flow_chart}
\end{figure}

For better understanding the difference among the multiple variants of TSSF based methods, CSP, and standard Riemannian methods, we summarize all the above steps into a flowchart (Fig.~\ref{Flow_chart}). After the prediction, the scoring metric chosen by us is the ROC-AUC (receiver operating characteristic - area under the curve) metric, and these scores are computed via five-fold cross-validation within each session of every data set.

After obtaining scores from different pipelines, the next step is to compare and analyze their performance statistically. In our work, two statistics, the $p$-value and the effect size, are adopted to compare the proposed TSSF against CSP as well as the full Riemannian approach. The $p$-value for the one-sided test is computed across sessions and subjects but within each data set, the null hypothesis of which is that the median accuracy of using one pipeline is not larger than using another pipeline. The effect size is measured by the standardized mean difference (SMD) between the accuracies of the two compared methods. Further details about these statistical tests can be found in \cite{jayaram2018moabb}.

\section{Results}
To comprehensively assess the performance of the proposed TSSF, three aspects are considered in this paper: the quality of the filtered feature, the interpretability revealed from associated spatial patterns, and the computational time. 

Of these three perspectives, feature quality is the only indicator which can be analyzed in a purely quantitative manner through the classification accuracy. Therefore, in this section, we exclusively analyze the performance of feature quality. Although it can be argued that the results of computational time can be analyzed in a quantitive way, i.e., by exhaustively comparing the simulation results of computational time, we more concerned with the theoretical, computational complexity analysis since the latter is more general than the simulation results. Therefore, interpretability and computational time are both left until the discussion, as the results are more qualitative and require more context to be properly interpreted. 

\begin{figure*}[htb!]
	\centering
	\begin{subfigure}{.78\linewidth}
		\includegraphics[width=\linewidth]{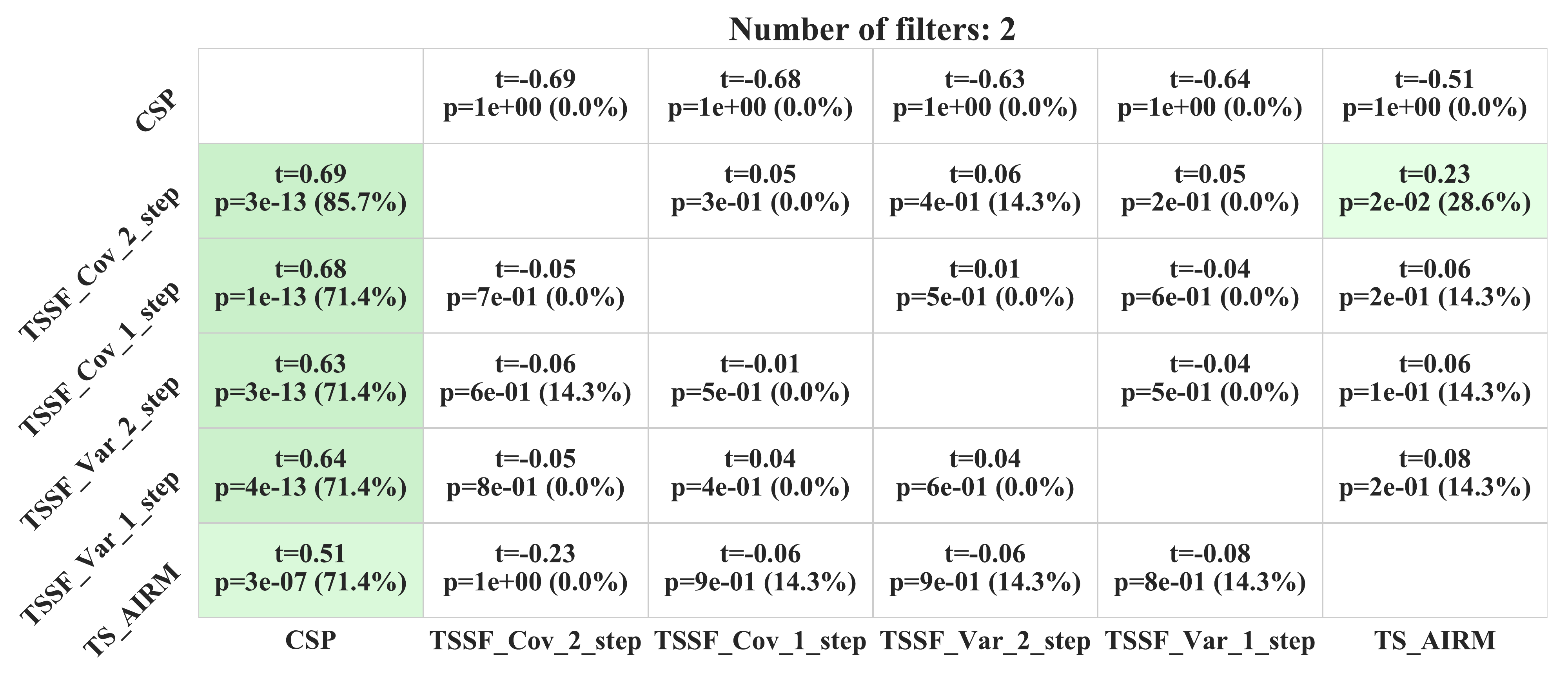}
		\caption{\centering Applying the first two filters}
		\label{Fig_FQ_summary_1}
	\end{subfigure}
	\begin{subfigure}{.78\linewidth}
		\includegraphics[width=\linewidth]{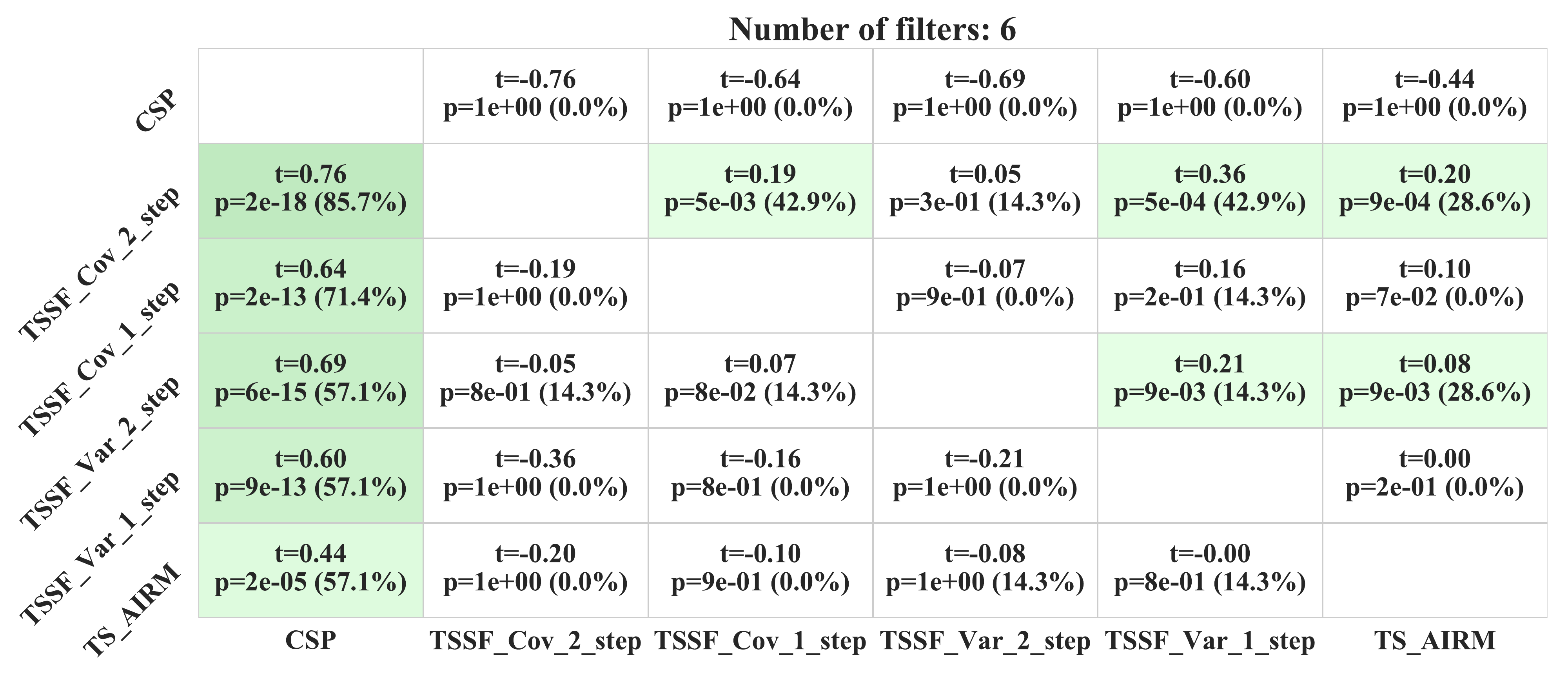}
		\caption{\centering Applying the first six filters}
		\label{Fig_FQ_summary_2}
	\end{subfigure}
	\begin{subfigure}{.78\linewidth}
		\includegraphics[width=\linewidth]{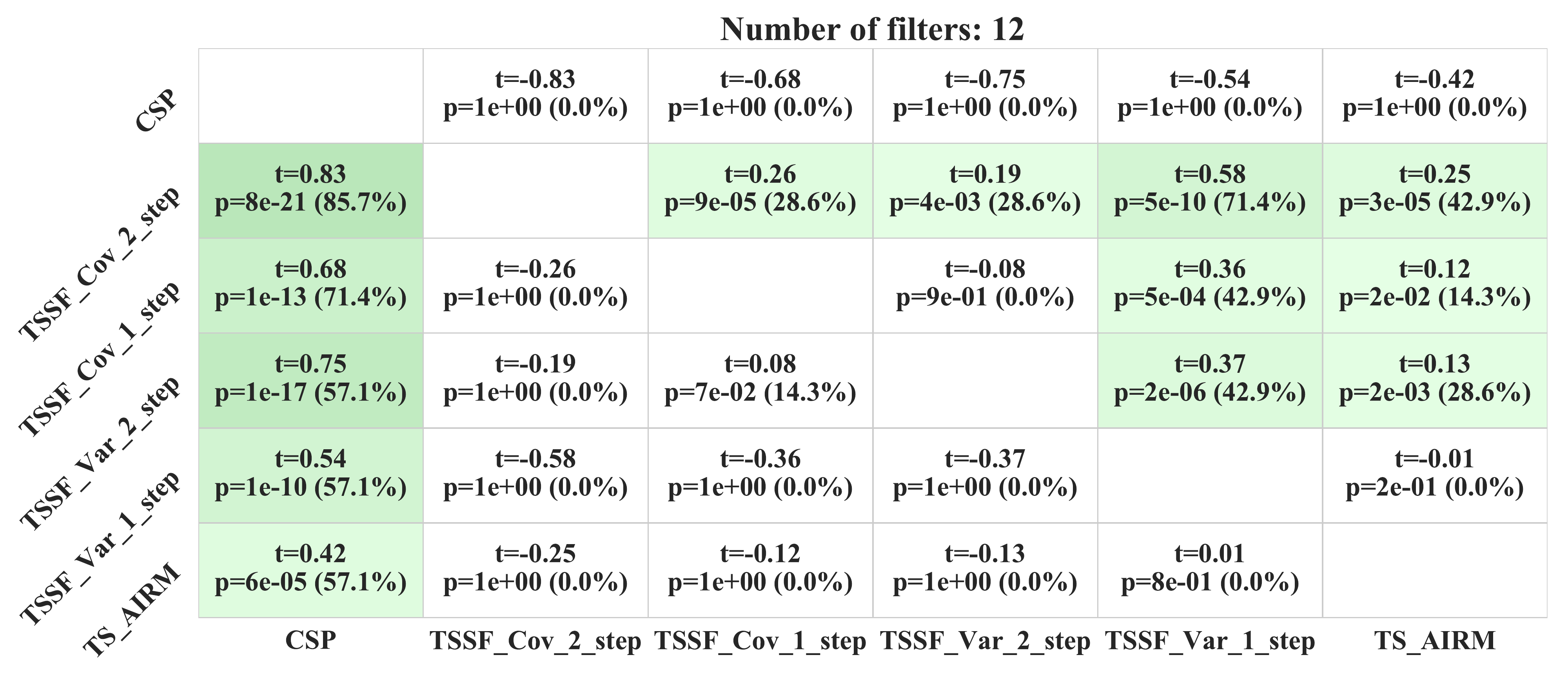}
		\caption{\centering Applying the first twelve filters}
		\label{Fig_FQ_summary_3}
	\end{subfigure}
	\caption{Statistical comparison of the classification accuracies from different pipelines. Parameters: effect size $t$ (standardized mean difference) and p-value $p$ are computed within each dataset. In each block, the statistics are computed based on the null-hypothesis that the median accuracy of row method is not larger than the column method. The green block means there exists an overall significant result across all datasets. The red block means there exist contradictory results, i.e., the overall one-tailed results show significance, but the effect size is not positive. Furthermore, the number in parentheses next to the p-values represents that the percentage of datasets in which significance is reported. The meaning of each label can be referred from Table.~\ref{Feat_table} and Fig.~\ref{Flow_chart}}.
	\label{Fig_FQ_summary}
\end{figure*}

As a typical indicator of feature quality, the classification accuracies are chosen to be compared as a way of assessing which features are most informative. In subsequent subsections, we begin with a comparison of all proposed classification pipelines over all the datasets, to see whether any of them consistently outperform the rest. The results are shown in Figure \ref{Fig_FQ_summary}. 

\subsection{Statistical performance across datasets}  \label{stat_analysis} 
We select three typical cases of applying spatial filters: two, six, or twelve spatial filters. By observing Fig.~\ref{Fig_FQ_summary_1} we can notice that even when only applying two filters, the p-value of comparison between all TSSF-based pipelines and \textit{CSP} highly significant, and the effect sizes are moderate. Moreover, in the comparisons with the full Riemannian method, the \textit{TSSF\_Cov\_2\_step} even significantly outperforms the full Riemannian method, albeit with a small effect size (0.23).

When increasing the filter number to 6, as shown in Fig.~\ref{Fig_FQ_summary_2}, the performance of all TSSF-based pipelines continues to surpass \textit{CSP}. Surprisingly, \textit{TSSF\_Var\_2\_step} also shows significantly better results than the full Riemannian method \textit{TS\_AIRM}, though again with a rather small effect size (0.08). In addition, performance begins to differ among the different TSSF-based pipelines. After increasing the number to 12 (Fig.~\ref{Fig_FQ_summary_3}), although one more TSSF-based pipeline significantly outperforms \textit{TS\_AIRM}, the differences among the TSSF-based pipelines also further enlarge.

Observing and comparing these figures from a macro perspective, we can discover several trends: First, CSP is constantly outperformed by all Riemannian based methods. Second, the performances of TSSF-based methods tend to differ from each other, only at large numbers of filters. Third, the difference in performance between one-step and two-steps methods also enlarges as the filter number increases. 

\subsection{Statistical performance within each data set}
In order to go into more detail on how the various algorithms perform, in Fig.~\ref{Fig_FQ_meta}, we show meta-analyses between some individual algorithms which describe the per-dataset performance. 

We first compare \textit{CSP} against \textit{TSSF\_Var\_2\_step}, as both fit the same secondary classifier. In order to see the benefit of the matrix logarithm-based features as compared to the filtered log-variance features, we further provide the comparison between \textit{TSSF\_Var\_2\_step} and \textit{TSSF\_Cov\_2\_step}. Lastly, we are interested in seeing how the one-step classifier works, and so we also include \textit{TSSF\_Var\_1\_step}. As some datasets only contain 20 or fewer channels, we choose to always use 6 filters. The results are as shown in Fig.~\ref{Fig_FQ_meta}.
\begin{figure}[H]
    \centering
    \begin{subfigure}{.91\linewidth}
        \includegraphics[width=\linewidth]{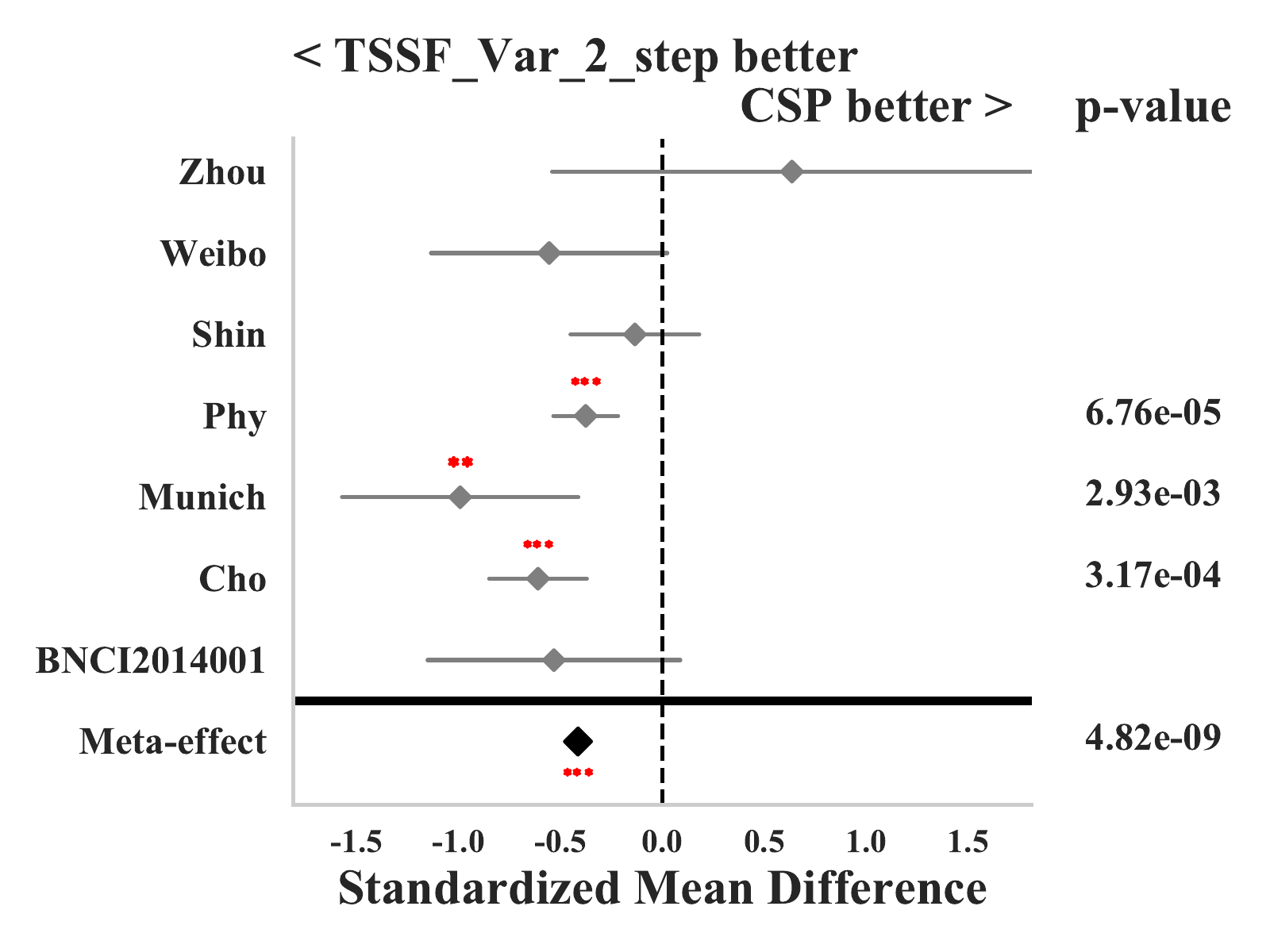}
        \caption{\centering Comparison of different spatial filter extraction methods}
        \label{Fig_FQ_meta_1}
    \end{subfigure}
\end{figure}

\begin{figure}[H]\ContinuedFloat
	\centering
	\begin{subfigure}{.95\linewidth}
		\includegraphics[width=\linewidth]{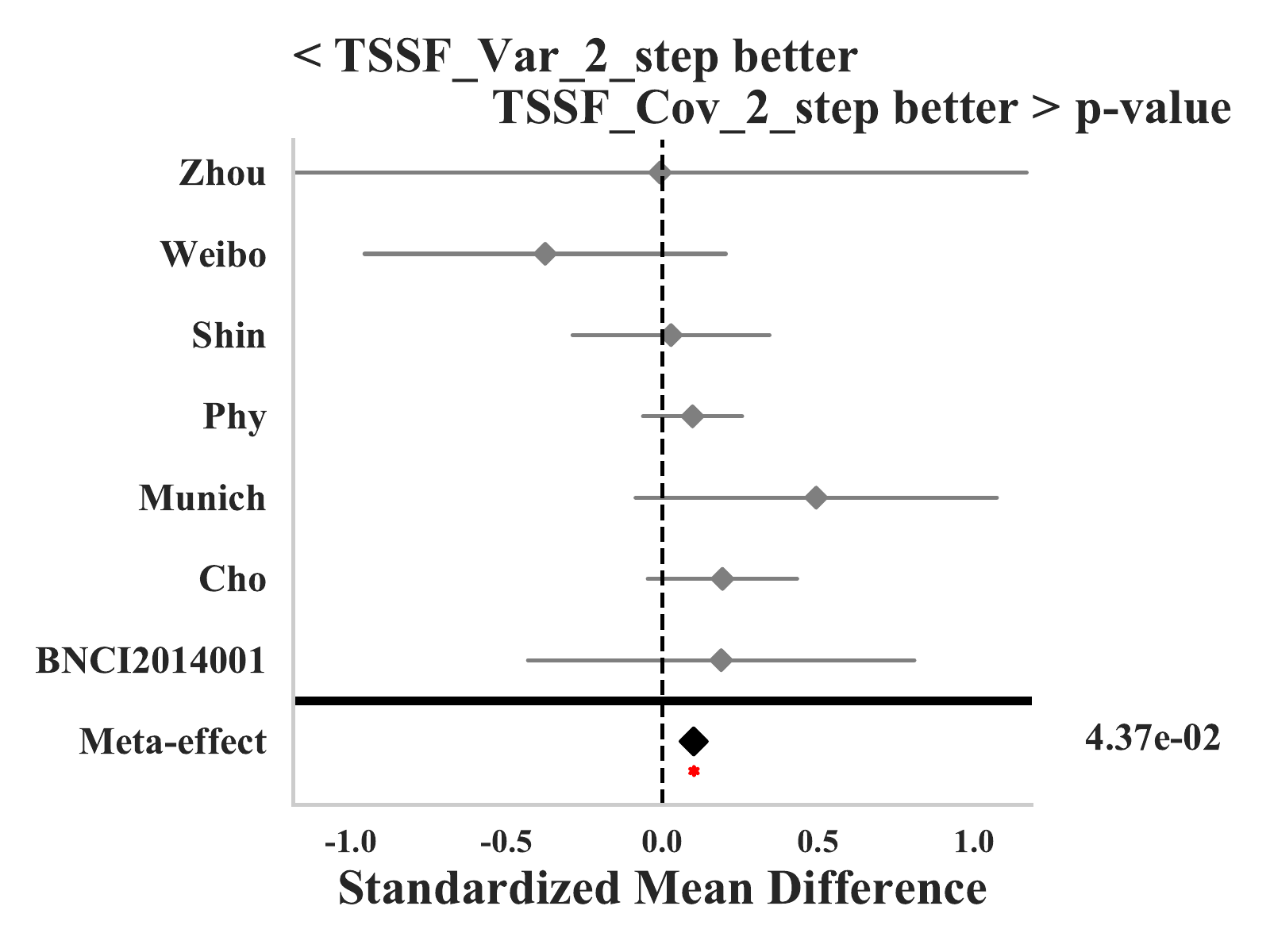}
		\caption{\centering Comparison of different feature types}
		\label{Fig_FQ_meta_2}
	\end{subfigure}
	\begin{subfigure}{.95\linewidth}
		\includegraphics[width=\linewidth]{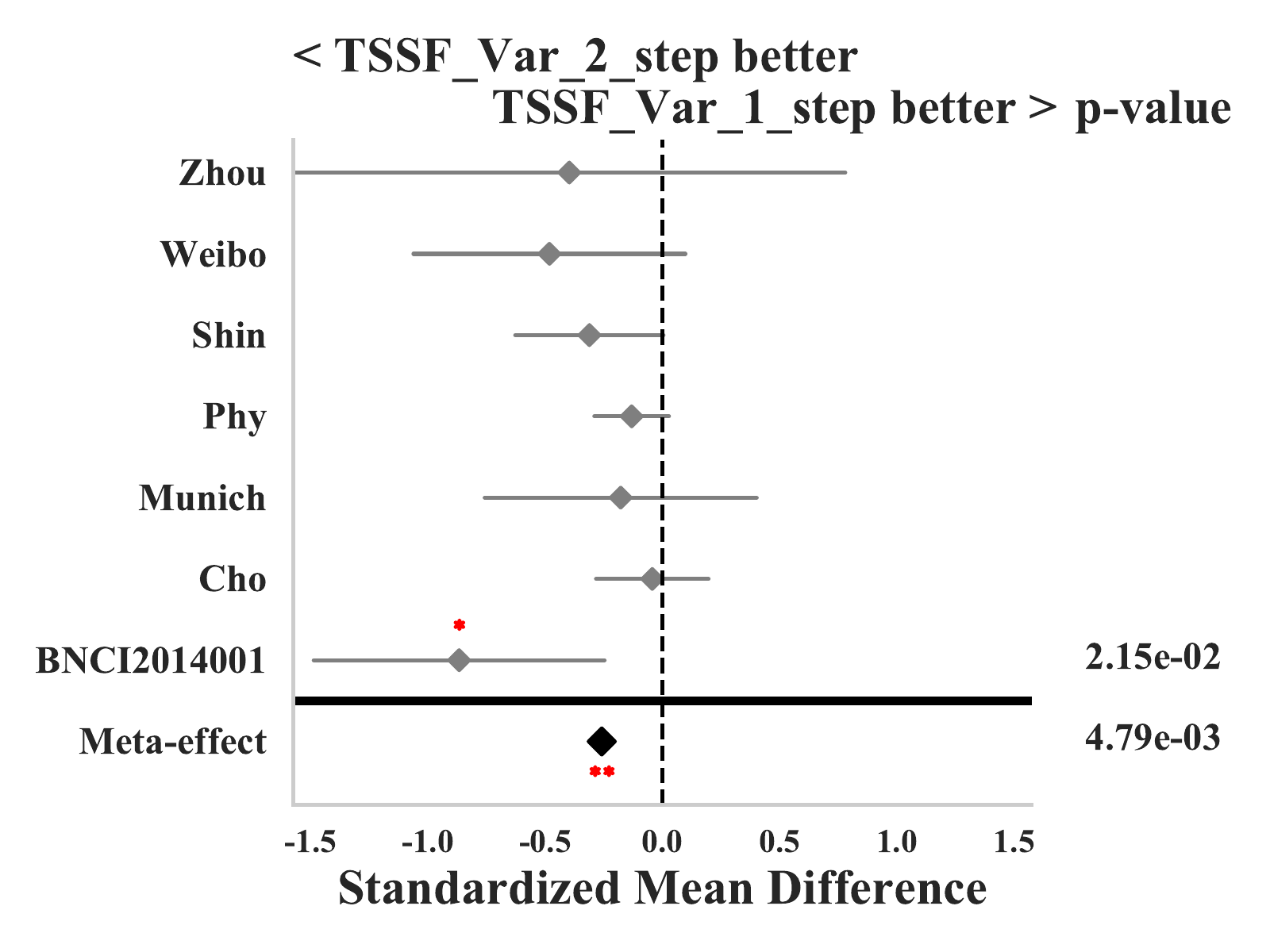}
		\caption{\centering Comparison of different classification methods}
		\label{Fig_FQ_meta_3}
	\end{subfigure}
	\caption{Meta analysis of accuracies using different pipelines with 6 filters. Parameters: p-value $p$ and SMD are computed within each dataset. Red \textcolor{red}{$^{\ast}$}, \textcolor{red}{$^{\ast\ast}$} and \textcolor{red}{$^{\ast\ast\ast}$} represent $p<0.05,0.01,0.001$ respectively. Grey diamonds signify the SMD, while grey bars show the confidence interval of the mean.}
	\label{Fig_FQ_meta}
\end{figure}

In the comparison of different spatial filter extraction method (Fig.~\ref{Fig_FQ_meta_1}), the TSSF-based method overwhelms CSP with only one exception. In addition, as shown in the Fig.~\ref{Fig_FQ_meta_2}, the features types does not seem to have a significant influence on the feature quality within each data set, even if the log-variance features is with dimension $dim = K=6$ and logarithm of covariance matrices features is $dim = \frac{K(K+1)}{2} = 21$. Although there exists a controversial fact that an overall significance appears in the comparison across datasets with $p=0.0436$, this chance-level p-value is not supported by a significant difference within any individual dataset. In the last sub-figure, Fig.~\ref{Fig_FQ_meta_3}, we can see that while two-step classification is significantly better than one-step classification across datasets, this is heavily influenced by only one data set. 

\begin{figure*}[t!]
	\centering
	\includegraphics[width=\linewidth]{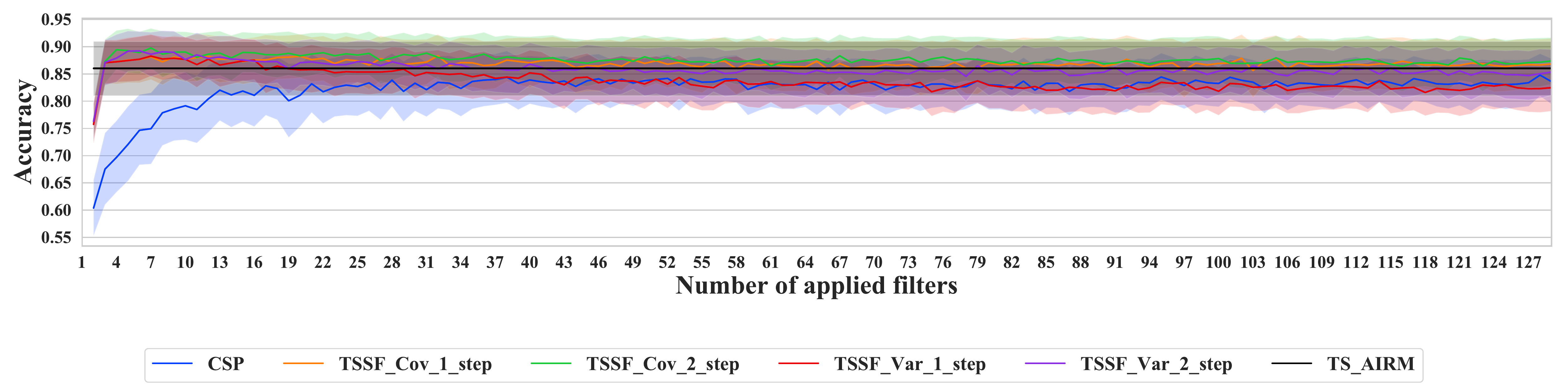}
	\caption {Classification accuracy w.r.t. the number of applied filters within Munich Motor Imagery data set and the accuracies are computed across all subjects and sessions. The central line is the mean accuracy and the error band shows confidence interval = $68\%$.}
	\label{Fig_FQ_acc}
\end{figure*}
\subsection{Performance w.r.t. the number of applied filters in single data set}
Last but not least, we look at how performance changes as a function of the number of applied filters. As a meta-analysis here results in an enormous number of statistical tests on not very much data, we focus on this section of the analysis on a single dataset. For better reflecting the relationship between accuracy and number of applied filters, we choose the data set \textit{Munich Motor Imagery}, which has the highest channels numbers (128). %
\begin{figure}[H]
	\centering
	\includegraphics[width=0.95\linewidth]{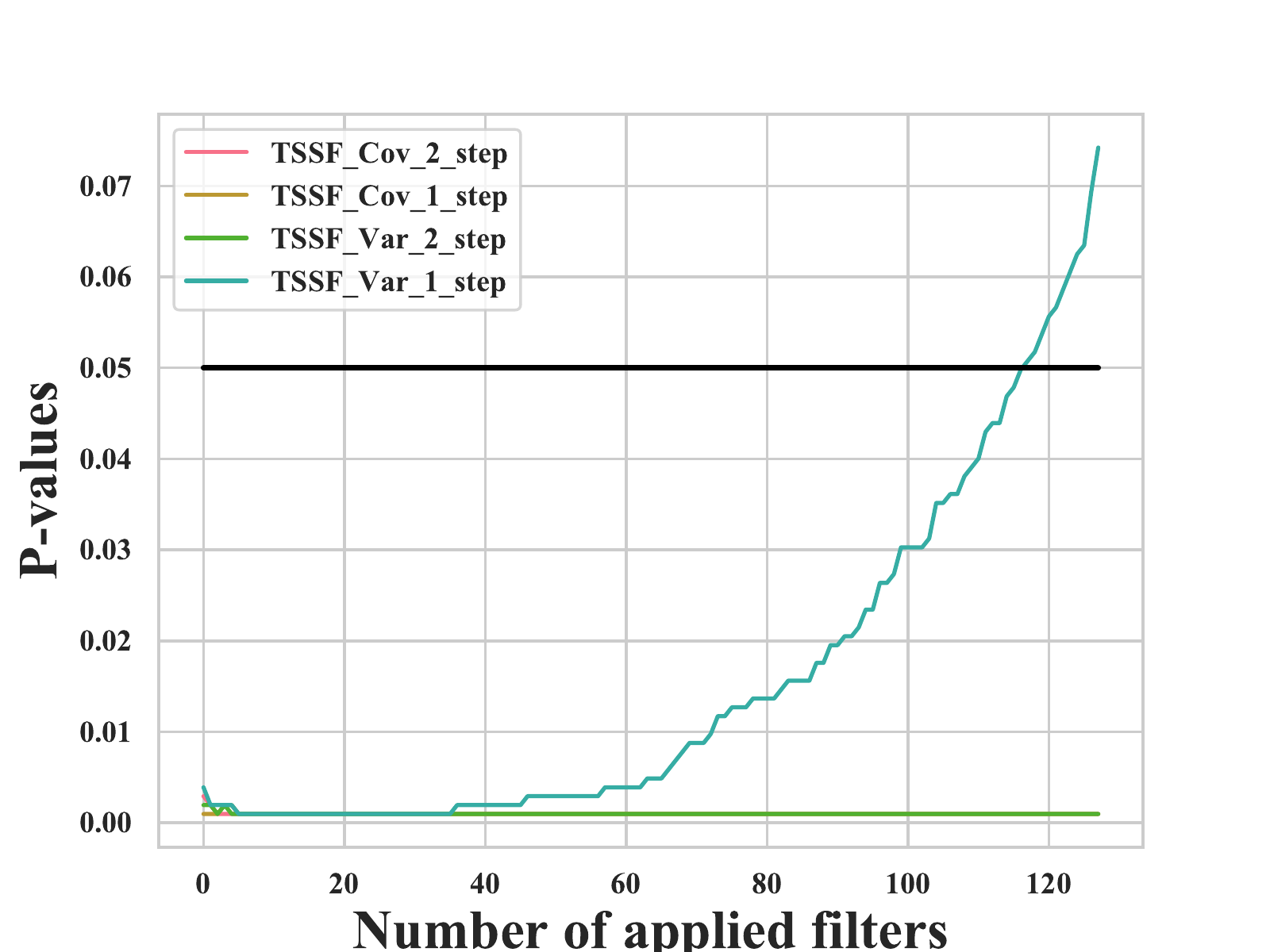}
	\caption{The p-values from the statistical test between all TSSF-based pipelines and CSP w.r.t. to the number of applied filters. The chosen data set is \textit{Munich Motor Imagery} and the null-hypothesis is that the median accuracy of TSSF-based methods is not larger than CSP. Significance threshold is set as 0.05, as indicated by the black straight line.}
	\label{Fig_CSP_pvalues}
\end{figure}

From Fig.~\ref{Fig_FQ_acc} we first notice that the accuracies of all TSSF-based features converge to the performance of the standard Riemannian method with merely four filters while CSP needs around 20 filters to reach a stable plateau. Second, except for \textit{TSSF\_Var\_1\_step}, all other TSSF-based methods constantly significantly outperform the \textit{CSP} whatever number of filters is used, as shown in Fig.~\ref{Fig_CSP_pvalues}. Lastly, for all log-variance based TSSF pipelines, their accuracy usually decreases when the number of filters continues to increase. Moreover, this fact can also be observed in other datasets, as indicated in Appendix \ref{all_ds_acc}.

In this section, we have comprehensively compared the quality of the features extracted from various ways, and confirmed two things: that Riemannian methods reliably outperform CSP, and that TSSF can approximate--and sometimes even outperform--standard Riemannian methods. As a spatial filtering method, however, the interpretability is always of the highest significance,  especially for online purposes, because it is the only way that we can know whether reasonable underlying neuronal sources are utilized. Moreover, the computational efficiency of the spatial filtering method is also vital because the online BCI system usually has a strict requirement for its computational complexity. Therefore, in the next section, we will further discuss these two aspects.

\section{Discussion}
We have shown that spatial filters can be extracted from linear functions in the Riemannian tangent space, and further that CSP can be seen as a special case of this general framework. This can be used to render Riemannian methods suitable for online use even in cases of over 100 channels. Moreover, we validate our approaches using over 220 subjects via an open-access toolkit \cite{jayaram2018moabb} and show that as far as classification accuracy is concerned, this method is statistically indistinguishable from the full tangent space approach on average, and in some cases can significantly improve on it. All in all, the proposed framework allows for the possibility of off-the-shelf algorithms made to work on vector data being used in the case of EEG data to generate spatial filters, eliminating the need for complicated optimization frameworks. 

One notable contribution of this paper is the proposal of one-step classification, which further reduces the computational time of the testing stage significantly, and so we begin our discussion there. Subsequently, we analyze the associated spatial patterns of both CSP and TSSF. Afterward, we discuss the signal sources as well as the robustness reflected from these patterns. We end our discussion with several suggestions regarding its usage and finally, a look towards the future work this result implies.

\subsection{One-step classification}

While one-step classification relies strongly on the assumption that the input points are roughly jointly diagonalizable, and hence that the proposed approximation holds, we have shown in practice that this appears to be the case for sufficiently small numbers of filters. What this suggests is that certain underlying sources can be extracted by static spatial filters, while others do not correspond to static eigenvectors of the covariance matrices. If few enough filters are chosen, the resulting classifier is very close to the tangent space function, but as more are added, the approximation quality degrades. This explains the results in Fig.~\ref{Fig_FQ_acc} in which the only classifier whose quality degraded as a function of filter number was the single-step log-variance classifier. 

Another major benefit to using one-step classification is that it is a better use of training data. Current spatial filtering-based approaches to classifications need to either re-use data for both spatial filter and classifier fitting, or partition training data into disjoint sets, which reduces the quality of both solutions. When the approximation holds, one-step classification is a much more data-efficient solution.

\subsection{One-step classification: Computational complexity analysis and simulation results }

\begin{figure}[H]
	\centering
	\includegraphics[width=\linewidth]{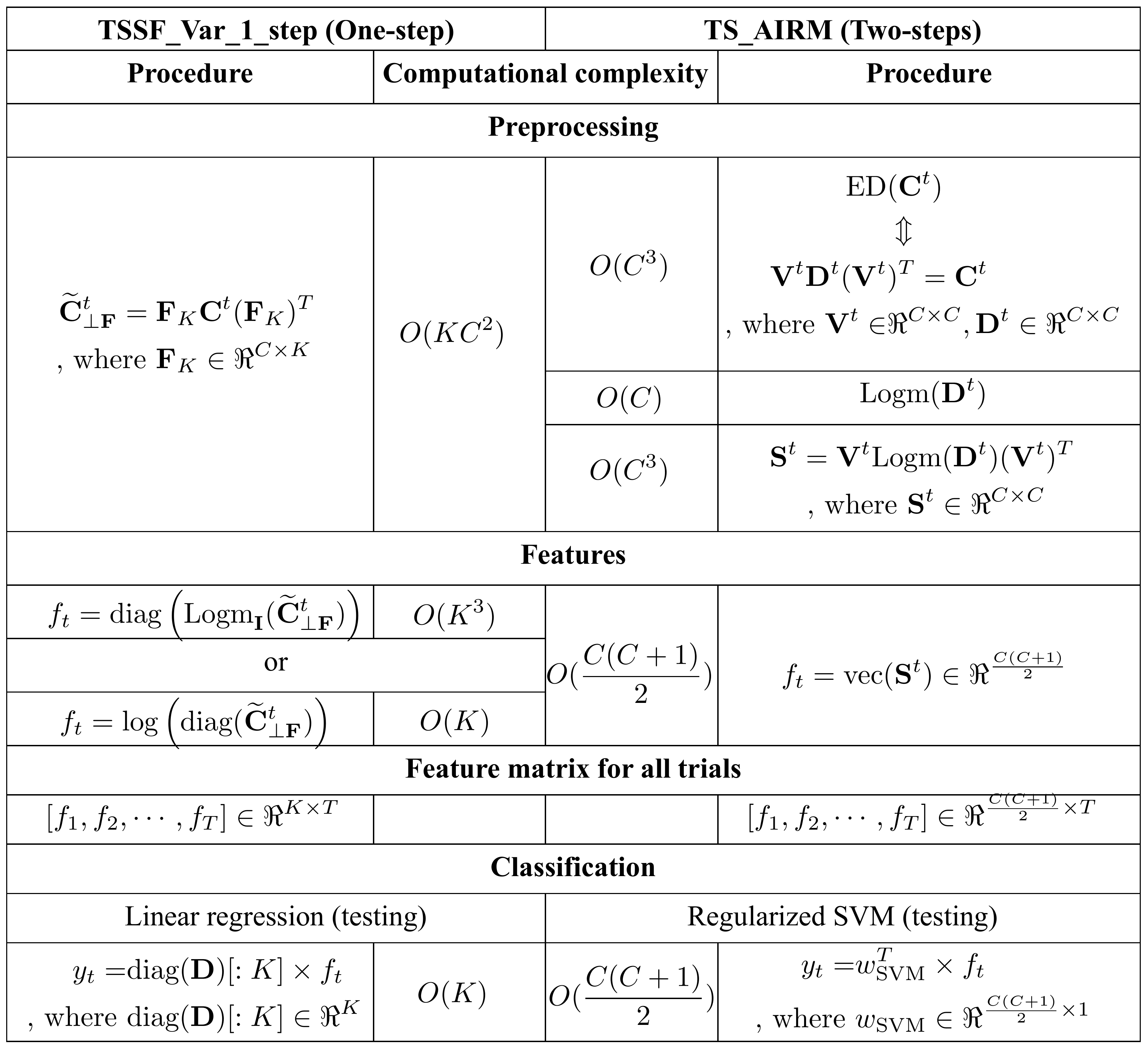}
	\caption{Theorectical computational complexity analysis and comparison between the examples of one-step and two-step classification in the testing stage. Note: all listed computational complexity are theoretical, and the practical complexity are usually smaller than the list value due to the adoption of specific algorithm. For instance, the matrix multiplication complexity is theoretically equal to $O(C^3)$ but usually between $O(C^{2.376})$ \cite{coppersmith1990matrix} and $O(C^3)$ in practice.}
	\label{Fig_Complexity}
\end{figure}

Considering that one major critique of Riemannian methods in online practice is their inability to scale to high numbers of channels, the computational complexity comparisons between the one-step classification framework and the full Riemannian tangent space method are provided from both the perspectives of theoretical analysis and simulation results.

\begin{figure}[htb!]
	\centering
	\includegraphics[width=\linewidth]{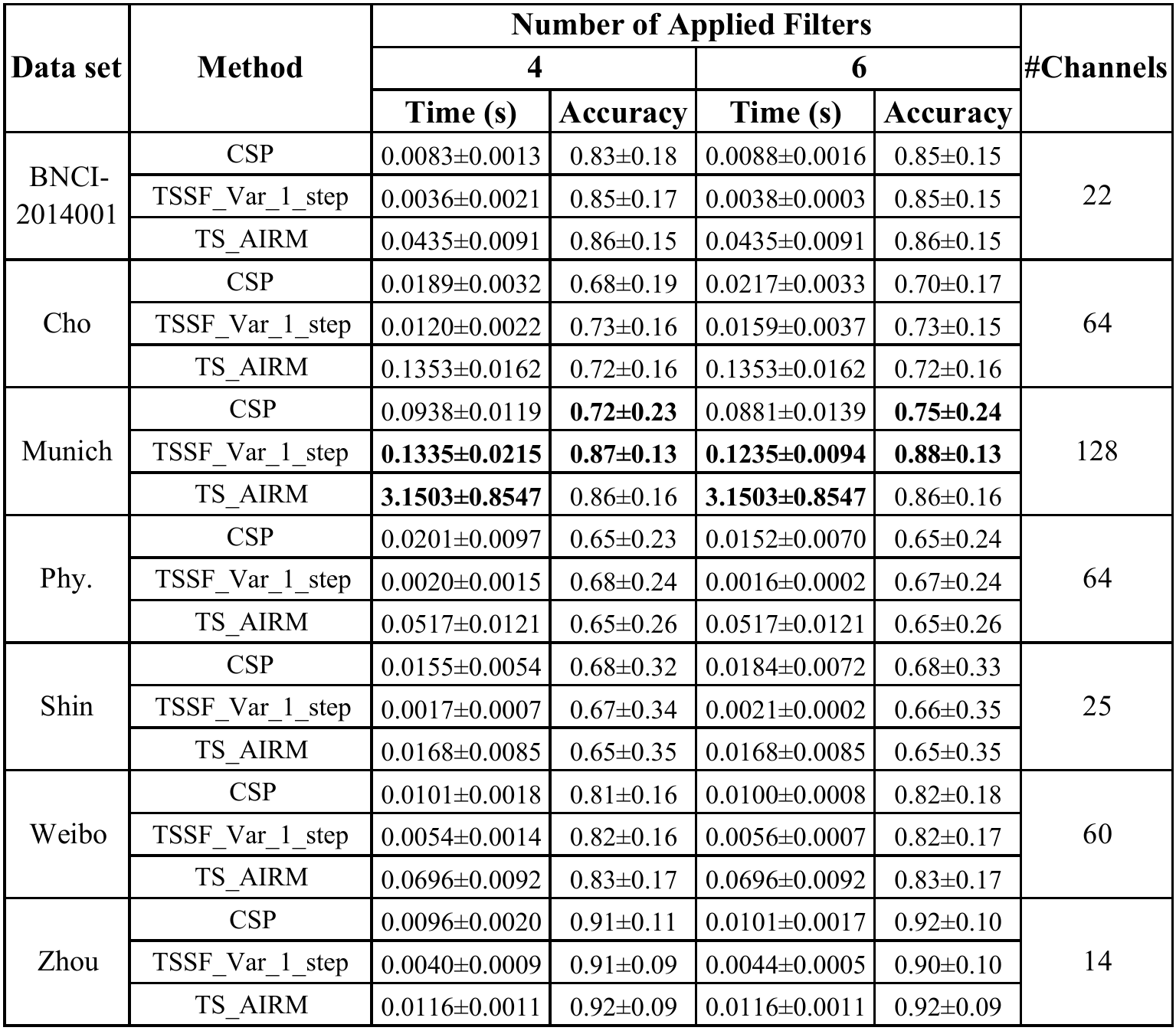}
	\caption{Comparison of classification accuracy and running time in the testing stage for three pipelines: \textit{CSP}, \textit{TSSF\_Var\_1\_step} and \textit{TS\_AIRM}. The values for both accuracy and time are with the format of the mean $\pm$ the standard deviation, which is computed across all sessions within each data set. The comparisons with the largest contrast are noted as bold. The above numbers are obtained from computers with 64GB RAM and an 8-core CPU.}
	\label{Fig_Time}
\end{figure}

For a better understanding of simulation results, we first start with the theoretical analysis. As seen from Fig.~\ref{Fig_Complexity}, standard Riemannian methods require operations with a computational complexity of either $O(C^3)$ or $O(\frac{C(C+1)}{2})$. For high numbers of channels, this can be difficult to do for real-time feedback, and to verify that we ran a theoretical runtime analysis using the \textit{Munich Motor Imagery} data set, as it has over 100 channels. The simulation results are shown in Fig.~\ref{Fig_Time}. As seen from these results, standard Riemannian methods are slower than both CSP and TSSF based methods. In particular, the full Riemannian methods is 25 times slower than TSSF based methods with similar performance when observing the results from the data set with 128 channels. As for the accuracy comparison, the superiority of TSSF based methods is already validated, as shown in Fig.~\ref{Fig_CSP_pvalues} in which \textit{TSSF\_Var\_1\_step} has an overall significantly better classification performance than \textit{CSP} when using four or six filters.

In summary, by adopting the one-classification, it is no longer impossible to enjoy the robustness and excellent performance of Riemannian methods in an online BCI system with high-dimensional data. One additional advantage is that, by employing fewer features, the model suffers less risk from overfitting.

\subsection{How robust is spatial filter order to artifacts?} \label{discuss_artifacts}

\begin{figure*}[h!]
	\centering
	\includegraphics[width=.9\linewidth]{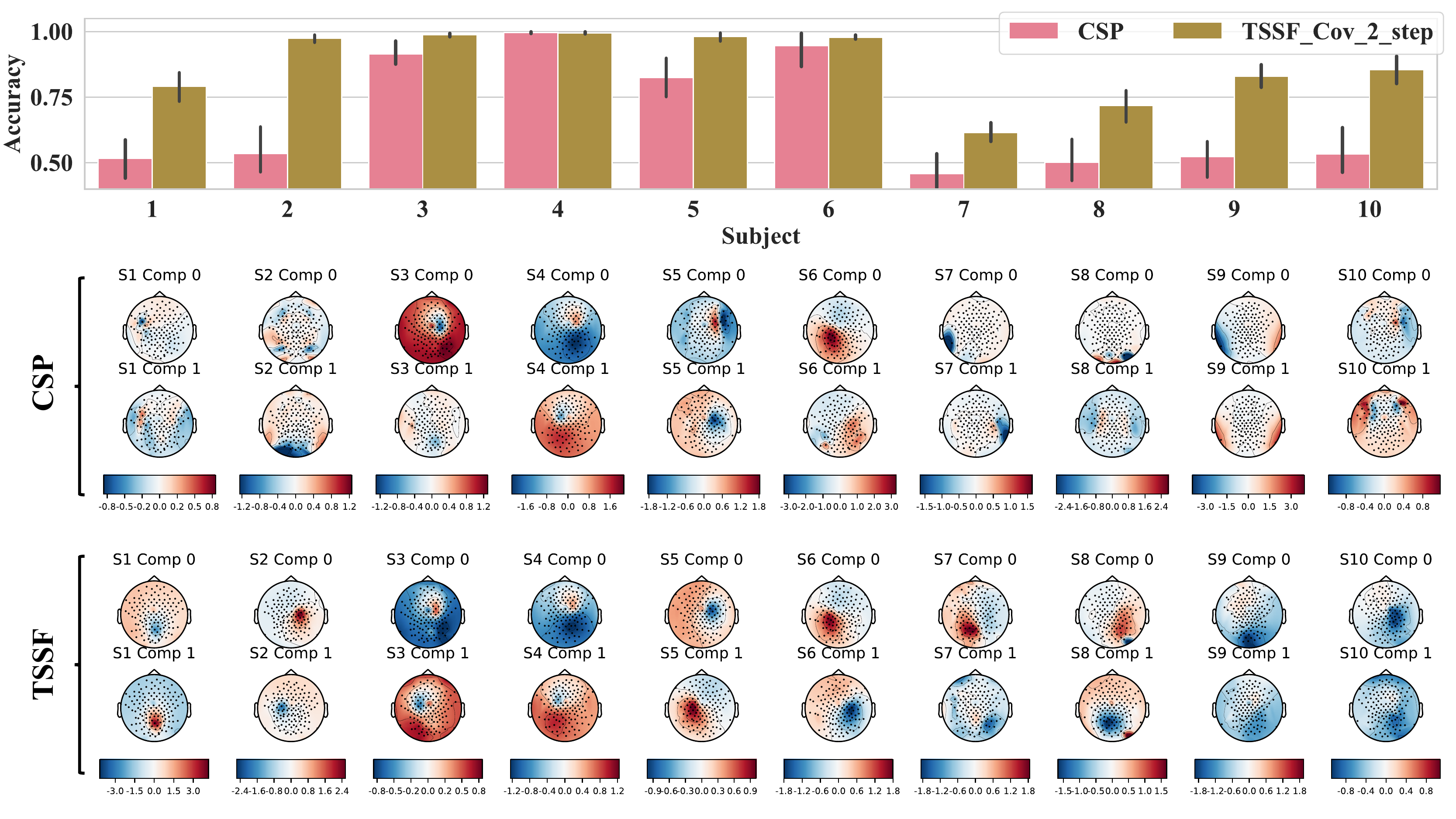}
	\caption{The classification accuracy and the associated spatial patterns of data set \textit{Munich Motor Imagery} when applying the first two spatial filters for all subjects.}
	\label{Fig_AD_pat}
\end{figure*}

Another important aspect of our work is the observation that this procedure allows one to easily validate the relevance of the features that a Riemannian classifier is using. By visualizing the spatial filters, it is easy to ensure that artifactual sources are not included in the classifier, which is of crucial importance when a BCI is used for neurofeedback.
As shown in Fig.~\ref{Fig_AD_pat}, in which only two filters are applied, the TSSF based methods are clearly better than CSP based, especially for S1, S2, S7, S8, S9, and S10. Correspondingly, we can notice that for these subjects, the patterns based on CSP look much more patchy than TSSF based. As for the subjects that both methods have tied performance, i.e., S4 and S6, their spatial patterns seem almost identical to each other.

\begin{figure*}[h!]
	\centering
	\begin{subfigure}{.9\linewidth}
		\includegraphics[width=\linewidth]{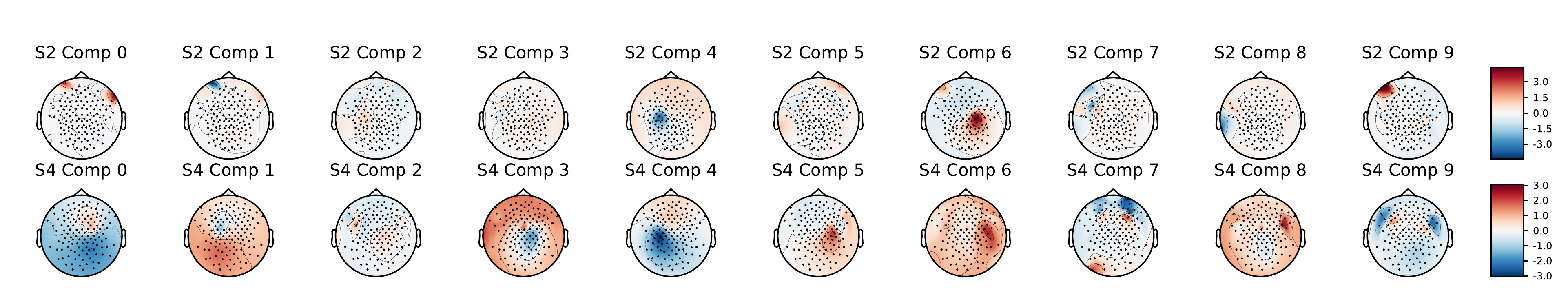}
		\caption{\centering Spatial patterns from CSP}
		\label{Fig_AD_10_pat_CSP}
	\end{subfigure}
	\begin{subfigure}{.9\linewidth}
		\includegraphics[width=\linewidth]{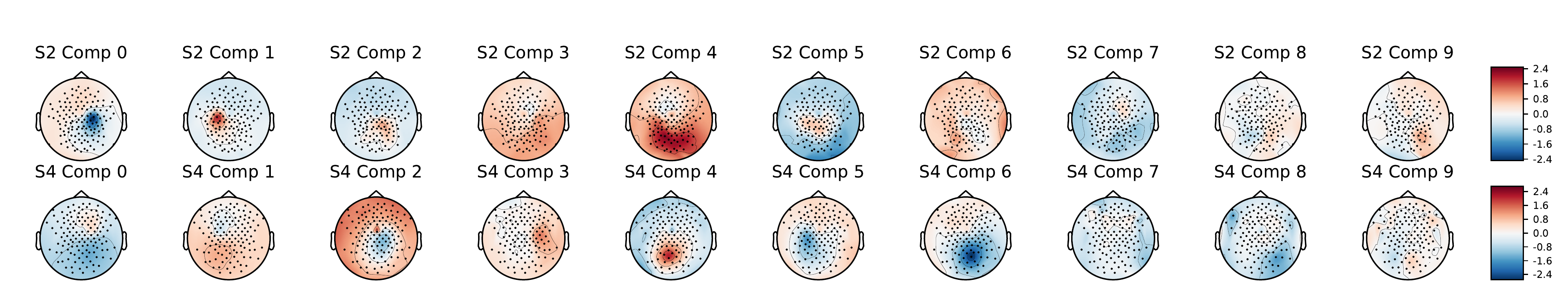}
		\caption{\centering  Spatial patterns from TSSF}
		\label{Fig_AD_10_pat_TSSF}
	\end{subfigure}
	\caption{The associated spatial patterns of data set \textit{Munich Motor Imagery} when applying the first ten spatial filters for S2 and S4. The major conclusion drawn from these two figures is that the TSSF based patterns present a better ordering comparing to the CSP which means the CSP tends to be affected by artifacts.}
	\label{Fig_AD_10_pat}
\end{figure*}

We would, however, like to better understand where and how components that are not brain-related enter the spatial filters in these two methods. Therefore, we increase the filter number to ten and compare the spatial patterns from two contrasting subjects in order to verify how the methods are robust to artifacts. The chosen subjects are S2 and S4 because as described in \cite{grosse2009beamforming}, they reflect two extremes of artifact contamination. In S2 55$\%$ of all trials are contaminated by artifacts while only 6.3$\%$ of all trials are affected by artifacts for S4. Therefore, their associated patterns are as shown in Fig.~\ref{Fig_AD_10_pat}.

Observing the Fig.~\ref{Fig_AD_10_pat}, S2 Comp 0 and S2 Comp 1 of the spatial pattern of TSSF seems similar to S2 Comp 4 and S2 Comp 6 of CSP. As for the rest CSP patterns of S2, most of them appear to be artifacts while for TSSF patterns of S2, only Comp 8 and Comp 9 look slightly patchy while the rest of the patterns show strong activity around the sensorimotor cortex. Moreover, in S4's patterns, from Comp 0 to Comp 6, the results of CSP and TSSF reflect similar neuronal sources with a slightly different order. However, when looking at the last three patterns of both filters, artifactual sources appear. Overall, the ordering in TSSF is much more informative than that in CSP, although in very low artifact scenarios they are similar. 

Therefore, we would like to conclude that the associated spatial patterns reflect more neurophysiologically explainable neural sources in TSSF. In contrast, CSP often gets distracted by artifacts, especially when processing data suffering plenty of contamination, e.g., S2 of data set \textit{Munich Motor Imagery}. When considering CSP as a simplified LDA based TSSF, this susceptibility to noise directions makes much more sense. When considering the patterns from the low-artifact subject S4, the patterns from both spatial filtering methods are almost identical with each other, in particular for the first several patterns. 

\subsection{How many filters lead to optimal performance?}

By enlarging the feature space, classification accuracy only increases when useful information is encoded within the additional features. For the case of spatial filtering, the most informative features are usually from the first several spatial filters and afterward, the features are no longer as informative as before, as indicated by the spatial patterns shown in Fig.~\ref{Fig_AD_10_pat}. Therefore, when applying only a few spatial filters, there always exists a positive relationship between the performance and number of applied filters. However, this positive relationship turns into a plateau when further increasing the filter numbers because the additional features are no longer as informative as before, and can even turn negative in cases of overfitting.

This initial positive relationship and the subsequent plateau always raises the question of the optimal number of spatial filters, as the optimal scenario for applying spatial filtering techniques is to use the least number of filters to achieve a given level of performance. We argue that TSSF reliably requires less spatial filters than CSP in order to achieve the same level of performance. The arguments are three-fold: 

First, CSP usually needs more filters than TSSF to reach the plateau in classification accuracy. For instance, as described in Fig.~\ref{Fig_FQ_acc}, TSSF based methods merely require 3 or 4 filters to reach 95\% of its best performance while CSP needs at least 20 components to reach the same level. Although in other datasets, the comparison is not as evident as in \textit{Munich Motor Imagery}, it is still clear that TSSF based methods converge faster as shown in Appendix \ref{all_ds_acc}. Second, the optimal minimum filter number for TSSF appears to be independent of the number of channels in the dataset, possibly reflecting a true biological set of sources conserved across all the data. For all seven datasets, after using six filters or even less, TSSF already reaches the level of its best performance, while for CSP, this number varies from 6 to 20 and shows a roughly positive correlation with the number of channels. Last but not the least, as the figures in Appendix \ref{all_ds_acc} indicate, the best classification performance of the TSSF methods is higher than of the CSP, and so even a suboptimal number of filters can compare with it.

For a given task, the truly activated neural area should be conserved across subjects and datasets as indicated by the neurophysiological knowledge, and so it is likely that TSSF more reliably extracts the neurophysiological signals independent of variables like setup and channel number. More specifically speaking, as shown in Fig.~\ref{Fig_AD_10_pat}, the associated spatial patterns when using ten filters of both methods looks pretty similar when processing high-quality data (only 6.3\% trials are contaminated for S4). However, for S2 in which 55\% trials are contaminated, CSP ranks the two most important neural sources as the fifth and seventh components, while TSSF ranks them at the top two components. %
From a machine learning point of view, this result is unsurprising if one considers CSP to be a simplified LDA, which does not take second-order information into account, and that our TSSF is done using an SVM in the original tangent space. However, what is surprising is the conservation of the number of necessary sources across the various datasets -- an SVM is sparse in terms of the number of support vectors it uses to build its projection, but these support vectors live in the tangent space. There is no obvious reason that the spatial filters derived from them should also be only informative in 6 out of 128 dimensions.

\subsection{The origin of the robustness and discriminability of Riemannian methods} \label{robustness}

As shown in Appendix \ref{all_ds_acc}, for all datasets, the accuracies of log-variance based methods usually first rise until reaching a peak and begin to drop after that as the filter number increases. Practically speaking, this is a confirmation of a well-known fact within spatial filtering: more filters are not always good. When approximating the Riemannian tangent space function, we find a similar trend: with very few filters, it is possible to outperform the full Riemannian tangent space classifier sometimes, as evidenced by Sections \ref{stat_analysis} and Fig.~\ref{Fig_FQ_summary}. However, in contrast to CSP, increasing the number of filters does not degrade performance as visibly. This robustness to the number of filters is yet another benefit of TSSF.

We might, however, ask why this particular feature exists. Intuitively, in the sense of accuracy, the superiority of covariance-based features should be thanks to the cross-channel covariances. This argument is indeed correct; however, not in the usual way. We normally expect cross-channel power terms to contribute significantly to the decision function on the tangent space such that the Riemannian methods are rather robust. However, this intuitive inference cannot explain the success of diagonal elements-based TSSF features, i.e., \textit{TSSF\_Cov\_1\_step}, \textit{TSSF\_Var\_1\_step} and \textit{TSSF\_Var\_2\_step}, which support the hypothesis that the off-diagonal terms (correlations between spatially filtered signals) are neglectable in comparison with the diagonal entries. 

After excluding the possibility that off-diagonal terms influence the robustness through direct contribution to the decision function, we might ask how they affect the matrix logarithm operation. Luckily, this can be investigated via comparing the accuracies from \textit{TS\_Cov\_2\_step}, \textit{TS\_Cov\_1\_step} and  \textit{TS\_Var\_1\_step}. We begin by reviewing the three compared feature spaces:
\begin{enumerate}
	\item \textit{TS\_Cov\_2\_step}: $\text{vec}\left(\text{Logm}_{\widetilde{\mathbf{C}}^{m}_{\perp \mathbf{F}}}(\widetilde{\mathbf{C}}^t_{\perp \mathbf{F}}) \right) \in \Re^{\frac{C(C+1)}{2} \times 1} $ \\
	\item \textit{TS\_Cov\_1\_step}: $\text{diag} \left( \text{Logm}_{\mathbf{I}}(\widetilde{\mathbf{C}}^t_{\perp \mathbf{F}})\right)  \in \Re^{C \times 1} $ \\
	\item \textit{TS\_Var\_1\_step}: $ \text{log} \left( \text{diag}(\widetilde{\mathbf{C}}^t_{\perp \mathbf{F}})\right)\in \Re^{C \times 1}  $
\end{enumerate}
where $t = 1, 2, \cdots, T.$

By observing the Fig.~\ref{Fig_FQ_acc} and the figures in Appendix \ref{Fig_acc_allds}, \textit{TS\_Cov\_2\_step} and \textit{TS\_Cov\_1\_step} do not significantly differ in classification accuracy. Although in Fig.~\ref{Fig_acc_allds}c and Fig.~\ref{Fig_acc_allds}d,  \textit{TS\_Cov\_2\_step} seems a bit better than \textit{TS\_Cov\_1\_step}, it is still impossible to assert that one pipeline is consistently superior to another, in particular considering the poor data quality of these two datasets. This phenomenon also partly validates our previous claim that the contribution from off-diagonal elements to the decision function is neglectable.

When comparing between the accuracy of \textit{TS\_Cov\_1\_step} and \textit{TS\_Var\_1\_step}, the differences arise solely from the distinct way the features are computed, as the linear regression coefficients are both the GED log-eigenvalues in both cases. The discrepancy between $\text{diag} \left( \text{Logm}_{\mathbf{I}}(\widetilde{\mathbf{C}}^t_{\perp \mathbf{F}})\right) $ and $ \text{log} \left( \text{diag}(\widetilde{\mathbf{C}}^t_{\perp \mathbf{F}})\right) $ comes from whether the $\widetilde{\mathbf{C}}^t_{\perp \mathbf{F}}$ is diagonal, i.e., the filtered cross-channel power terms are all with zero or not. If they are, both features are equivalent to each other. If not, $ \text{log} \left( \text{diag}(\widetilde{\mathbf{C}}^t_{\perp \mathbf{F}})\right) $  is the approximation of $\text{diag} \left( \text{Logm}_{\mathbf{I}}(\widetilde{\mathbf{C}}^t_{\perp \mathbf{F}})\right) $ and the approximation error depends on how close the diagonal elements of $\widetilde{\mathbf{C}}^t_{\perp \mathbf{F}}$ are to its eigenvalues. Therefore, combined with the Gershgorin circle theorem \cite{gershgorin1931uber}, we know that the differences in classification accuracy and robustness between the two features are related to the filtered cross-channel power terms. In other words, the smaller the cross-channel power of $\widetilde{\mathbf{C}}^t_{\perp \mathbf{F}}$ are, the more discriminative and robust the log-variance based features $ \text{log} \left( \text{diag}(\widetilde{\mathbf{C}}^t_{\perp \mathbf{F}})\right) $ will be.

In conclusion, the off-diagonal terms influence the robustness and discriminability of Riemannian methods via affecting the logarithm operation, instead of via a direct contribution to the decision function. In other words, if the correlations between filtered signals are non-zero, then source power is not equal to the log-variance, because the eigenvectors of $\widetilde{\mathbf{C}}^t_{\perp \mathbf{F}}$ are no longer the standard basis.

\subsection{Suggestions for the usage of TSSF}
After exhaustively benchmarking the TSSF based methods against conventional algorithms, we provide several suggestions to the reader who would like to use the TSSF method:

\begin{enumerate}
	\item \textit{Use the empirical covariance estimator when possible}: Many Riemannian methods recommend regularized covariance estimators such as the Ledoit-Wolfe estimator. However, since diagonal loading cannot be added during online use (as the spatially filtered variances are used), high regularization runs the risk of degrading the approximation. 
	
	\item  \textit{Choose the Riemannian metric carefully}: Most of the theoretical analysis is based on the assumption that the affine-invariant Riemannian metric is utilized. A similar property remains to be validated for other Riemannian metrics. 
	
	\item  \textit{Choice of features}: as discussed before, there are three types of features to be adopted, $\text{vec}\left(\text{Logm}_{\widetilde{\mathbf{C}}^{m}_{\perp \mathbf{F}}}(\widetilde{\mathbf{C}}^t_{\perp \mathbf{F}}) \right) $, $\text{diag} \left( \text{Logm}_{\mathbf{I}}(\widetilde{\mathbf{C}}^t_{\perp \mathbf{F}})\right) $ and $ \text{log} \left( \text{diag}(\widetilde{\mathbf{C}}^t_{\perp \mathbf{F}})\right) $. As computational complexity of the feature decreases, the signal to noise ratio is affected negatively. Therefore, the concrete choice of features highly depends on the experimental environment, e.g., number of channels, sufficient computing sources, etc. But based on our experience, $\text{diag} \left( \text{Logm}_{\mathbf{I}}(\widetilde{\mathbf{C}}^t_{\perp \mathbf{F}})\right) $  is a good candidate for a task which demands high accuracy, while $ \text{log} \left( \text{diag}(\widetilde{\mathbf{C}}^t_{\perp \mathbf{F}})\right) $ might be a better choice for a strict real-time requirement.
\end{enumerate}

\subsection{Future Work}
This paper covers the fundamental concept and proof of spatial filtering via the tangent space. However, there are still many interesting directions worthy of being explored later on. We will discuss these possible directions from two levels, the extension of the scientific idea and the extension of these proposed algorithms:

\begin{enumerate}
	
	\item \textit{Unsupervised dimensionality reduction and multi-class TSSF}: TSSF based methods usually perform rather well with few components, which implies there exists an optimal subspace of low dimensionality where the brain projects. In this paper, this optimal subspace is found in a supervised fashion. Whether we can leverage this idea to find an unsupervised dimensionality reduction will be an interesting open question.  Moreover, it will also be fascinating to further investigate whether the TSSF can be applied to multi-class classification considering the potential of TSSF in practical application.
	
	\item \textit{TSSF in comodulation manner}: The proposed TSSF is currently extracted in a regression-like manner because it is derived from the inner product between a weight vector and a data vector. Whether we can also leverage continuous information encoded within the target variables, just like the source power comodulation (SPoC) method \cite{dahne2014spoc}, will also be very worthwhile to explore.
	
	\item \textit{Other choices of the first classifier}: Although the SVM based TSSF methods have indeed achieved a satisfying performance as presented in this paper, no matter from the perspective of classification accuracy or the interpretability, we will not assert that regularized SVM represents a global optimum. On the contrary, it will be very interesting to explore the influence brought by the selection of the classification algorithm on the tangent space.
	
	\item \textit{Multiple frequency bands}: In this paper, the features are extracted from the joint $\mu$ and $\beta$ band, i.e., from 8Hz to 32Hz. Hence, it remains a mystery whether the ampler information induced by the filter bank TSSF will outperform current TSSF based methods, just like the enhancement of filter bank CSP \cite{ang2008filter} comparing to CSP.

\end{enumerate}

\section{Conclusion}
Thanks to its impressive performance, the Riemannian manifold classification framework has seen an upsurge in interest in recent years. Historically, it has been hampered by various issues, namely that Riemannian methods scale poorly to high-density setups and are somewhat difficult to introspect.

To tackle these obstacles, we have proposed a set of methods based on the combination of spatial filtering techniques with Riemannian methods, because the former possess nice properties which Riemannian methods are lacking, such as low dimensionality and the visualization of signal sources (or associated spatial patterns). In order to further simplify the computation of the proposed idea, we have proved the rationality behind several variants of Riemannian features based on the approximation of the decision function on the tangent space. Moreover, we have also put forward one-step classification in order to simultaneously find a classifier and spatial filters. We hope that this work will allow for the expansion of Riemannian geometry-based methods into more BCI applications, and that it might spur further development in both application and theory for this sort of interface.

\begin{appendices}
	
\section{Eigenvector invariance after logarithm mapping} \label{Appendix A}
 
The generalized eigenvalue decomposition (GED) is defined as below:
\begin{equation}
\label{Eq:appendix-A-1}
\begin{aligned}
\text{GED}(\mathbf{C}^w, \mathbf{C}^m) \Leftrightarrow  \mathbf{C}^w\mathbf{F} = \mathbf{C}^{m}\mathbf{F} \mathbf{D} , 
\end{aligned}
\end{equation}
where $\mathbf{F}$ represents the generalized eigenvectors and  $\mathbf{D}$ is the generalized eigenvalue matrix. For the convenience, we denote the solutions of this GED problem as:
\begin{equation}
\label{Eq:appendix-A-2}
\begin{aligned}
\mathbf{F}, \mathbf{D}  = \text{GED}(\mathbf{C}^w, \mathbf{C}^m)
\end{aligned}
\end{equation}

Since $ \mathbf{C}^m$ is a SPD matrix which means it is invertible, equation (\ref{Eq:appendix-A-1}) could be expressed as below by left multiplying $ \mathbf{C}^{-m}$ for both sides:
\begin{equation}
\label{Eq:appendix-A-4}
\begin{aligned}
\text{GED}(\mathbf{C}^w, \mathbf{C}^m) & \Leftrightarrow  \mathbf{C}^{-m} \mathbf{C}^w\mathbf{F} =\mathbf{F} \mathbf{D} \\ & \Leftrightarrow  \mathbf{F}, \mathbf{D}  = \text{ED}(\mathbf{C}^{-m} \mathbf{C}^w)  \\
\end{aligned}
\end{equation}

By leveraging the SPD property, we can decompose the $\mathbf{C}^{-m}$ into $ \mathbf{C}^{-\frac{m}{2}}   \mathbf{C}^{-\frac{m}{2}} $, which is as proved below:
\begin{equation}
\label{Eq:appendix-A-5.pre}
\begin{aligned}
\mathbf{C}^{-m} & = \mathbf{V}^{m} \left( \mathbf{D}^{m}\right)^{-1}  \left( \mathbf{V}^{m}\right)\tran \\
& \overset{\left( \mathbf{D}^{m}\right)^{-1} > 0}{=}  \mathbf{V}^{m} \left( \mathbf{D}^{m}\right)^{-\frac{1}{2}}  \mathbf{I} \left( \mathbf{D}^{m}\right)^{-\frac{1}{2}}    \left( \mathbf{V}^{m}\right)\tran  \\
& \overset{ \left( \mathbf{V}^{m}\right)\tran  \mathbf{V}^{m}  = \mathbf{I}  }{=} \mathbf{V}^{m} \left( \mathbf{D}^{m}\right)^{-\frac{1}{2}}   \left( \mathbf{V}^{m}\right)\tran  \mathbf{V}^{m}  \left( \mathbf{D}^{m}\right)^{-\frac{1}{2}}    \left( \mathbf{V}^{m}\right)\tran \\
& = \mathbf{C}^{-\frac{m}{2}}   \mathbf{C}^{-\frac{m}{2}} \\
\end{aligned}
\end{equation}
Similarly,  $ \mathbf{C}^{+\frac{m}{2}}   \mathbf{C}^{-\frac{m}{2}} $  can be proved as equivalent to identity matrix $\mathbf{I}$. Hence, based on these ingredients derived by SPD property, equation (\ref{Eq:appendix-A-4}) is further written as:
\begin{equation}
\label{Eq:appendix-A-5}
\begin{aligned}
& \mathbf{F}, \mathbf{D}  = \text{ED}(\mathbf{C}^{-m} \mathbf{C}^w)  \\ &\Leftrightarrow \mathbf{C}^{-m} \mathbf{C}^w\mathbf{F} =\mathbf{F} \mathbf{D} \\
&  \Leftrightarrow  \mathbf{C}^{-\frac{m}{2}}  \mathbf{C}^{-\frac{m}{2}} \mathbf{C}^w  \mathbf{C}^{-\frac{m}{2}}  \mathbf{C}^{\frac{m}{2}} \mathbf{F} = \mathbf{C}^{-\frac{m}{2}} \mathbf{C}^{\frac{m}{2}}\mathbf{F} \mathbf{D} \\
&  \Leftrightarrow  \mathbf{C}^{-\frac{m}{2}} \mathbf{C}^w  \mathbf{C}^{-\frac{m}{2}}  \mathbf{C}^{\frac{m}{2}} \mathbf{F} =  \mathbf{C}^{\frac{m}{2}}\mathbf{F} \mathbf{D} \\
&  \Leftrightarrow  \mathbf{C}^{-\frac{m}{2}} \mathbf{C}^w  \mathbf{C}^{-\frac{m}{2}} \mathbf{V} =  \mathbf{V}\mathbf{D} \text{, where } \mathbf{V} = \mathbf{C}^{\frac{m}{2}} \mathbf{F}   \\
&  \Leftrightarrow \mathbf{V}, \mathbf{D}  = \text{ED}(   \mathbf{C}^{-\frac{m}{2}} \mathbf{C}^w  \mathbf{C}^{-\frac{m}{2}} )  
\end{aligned}
\end{equation}

Note that as $\mathbf{C}^{-m}\mathbf{C}^w$ is not symmetric, its eigenvalue decomposition is not constrained to be orthogonal. Moreover, since $  \mathbf{C}^{-\frac{m}{2}} \mathbf{C}^w  \mathbf{C}^{-\frac{m}{2}}$ is symmetric, the eigenvectors are orthogonal basis, which could be formulated as below:
\begin{equation}
\label{Eq:appendix-A-5-}
\begin{aligned}
 \mathbf{V}\tran \mathbf{V} &= \mathbf{I} \\
 \mathbf{F}\tran \mathbf{F} &\neq \mathbf{I} \\
  \mathbf{F}\tran \mathbf{C}^{m}\mathbf{F} &= \mathbf{I} \\
\end{aligned}
\end{equation}

On the other hand, $ \text{ED}(   \mathbf{C}^{-\frac{m}{2}} \mathbf{S}^w  \mathbf{C}^{-\frac{m}{2}} )  $ can be solved as:

\begin{equation}
\label{Eq:appendix-A-6}
\begin{aligned}
& \text{ED}(   \mathbf{C}^{-\frac{m}{2}} \mathbf{S}^w  \mathbf{C}^{-\frac{m}{2}} ) \\
& \Leftrightarrow \text{ED}(   \mathbf{C}^{-\frac{m}{2}} \left[  \mathbf{C}^{\frac{m}{2}}\text{Logm}\left( \mathbf{C}^{-\frac{m}{2}}  \mathbf{C}^w  \mathbf{C}^{-\frac{m}{2}} \right)  \mathbf{C}^{\frac{m}{2}}\right]  \mathbf{C}^{-\frac{m}{2}} )\\
& \Leftrightarrow \text{ED}( \text{Logm}\left( \mathbf{C}^{-\frac{m}{2}}  \mathbf{C}^w  \mathbf{C}^{-\frac{m}{2}} \right)  ) \\
& \Leftrightarrow \text{ED}\left(  \text{Logm}\left(\mathbf{V} \mathbf{D} \mathbf{V}\tran \right)  \right)  \\
& \Leftrightarrow \text{ED}\left( \mathbf{V}  \text{Logm}\left(  \mathbf{D}  \right) \mathbf{V}\tran \right) 
\\
& \Rightarrow \mathbf{V},  \text{Logm}\left(  \mathbf{D}  \right) = \text{ED}\left( \mathbf{V} \text{Logm}\left(  \mathbf{D}  \right) \mathbf{V}\tran \right) \\
& \Leftrightarrow \mathbf{V},  \text{Logm}\left(  \mathbf{D}  \right) =  \text{ED}(   \mathbf{C}^{-\frac{m}{2}} \mathbf{S}^w  \mathbf{C}^{-\frac{m}{2}} )
\end{aligned}
\end{equation}

Based on the results shown in equation (\ref{Eq:appendix-A-5}) and (\ref{Eq:appendix-A-6}), we know:
\begin{equation}
\label{Eq:appendix-A-7}
\begin{aligned}
& \mathbf{V},  \text{Logm}\left(  \mathbf{D}  \right) =  \text{ED}(   \mathbf{C}^{-\frac{m}{2}} \mathbf{S}^w  \mathbf{C}^{-\frac{m}{2}} ) \\
& \Leftrightarrow \mathbf{F}, \text{Logm}\left(  \mathbf{D}  \right)  = \text{ED}(\mathbf{C}^{-m} \mathbf{S}^w) \\
& \Leftrightarrow \mathbf{F}, \text{Logm}\left(  \mathbf{D}  \right)  = \text{GED}(\mathbf{S}^w, \mathbf{C}^{m} ) 
\end{aligned}
\end{equation}

By combining the results of equation (\ref{Eq:appendix-A-2}) and equation (\ref{Eq:appendix-A-7}), we have:
\begin{equation}
\label{Eq:appendix-A-8}
\begin{aligned}
\mathbf{F},  \mathbf{D} &= \text{GED}(\mathbf{C}^w, \mathbf{C}^{m} )  \\
\mathbf{F}, \text{Logm}\left(  \mathbf{D}  \right)  &= \text{GED}(\mathbf{S}^w, \mathbf{C}^{m} ) 
\end{aligned}
\end{equation}
Therefore, as a conclusion, the eigenvectors of GED problem keep invariant after logarithm mapping as long as the projecting matrix is not reference matrix.

\section{Proof of Lemmas}

\subsection{Lemma 1} \label{GED_approx_lemma1}
\textbf{\textit{Lemma 1: Invariant inner product between tangent vectors after affine transformation}}

\begin{equation}
\begin{aligned}
\label{Eq:appendix-B-Lemma-0}
<\mathbf{S}_1, \mathbf{S}_2>\mid_{\mathbf{C}_{\text{ref}}} &=<\mathbf{C}_{\text{ref}}^{-\frac{1}{2}}\mathbf{S}_1\mathbf{C}_{\text{ref}}^{-\frac{1}{2}}, \mathbf{C}_{\text{ref}}^{-\frac{1}{2}}\mathbf{S}_2\mathbf{C}_{\text{ref}}^{-\frac{1}{2}}>\mid_{\mathbf{I}}\\
&=\Tr{\mathbf{C}_{\text{ref}}^{-\frac{1}{2}}\mathbf{S}_1\mathbf{C}_{\text{ref}}^{-\frac{1}{2}} \cdot \mathbf{C}_{\text{ref}}^{-\frac{1}{2}}\mathbf{S}_2\mathbf{C}_{\text{ref}}^{-\frac{1}{2}}}, \\
\end{aligned}
\end{equation}

\textbf{\textit{Proof:}} $<\mathbf{S}_1, \mathbf{S}_2>\mid_{\mathbf{C}_{\text{ref}}} $ represents the inner product between tangent vector $\mathbf{S}_1$ and $\mathbf{S}_2$. Besides, these tangent vectors are projected on the tangent space based on the reference point $\mathbf{C}_{\text{ref}}$. For more details about the proof, please refer to  section 4 of \cite{pennec1998uniform} and section 3 of \cite{pennec2006riemannian}.

\textbf{Q.E.D.}

\subsection{Lemma 2} \label{GED_approx_lemma2}

\textbf{\textit{Lemma 2: Equivalence of logarithm mapping}}
\begin{equation}
\begin{aligned}
	\label{Eq:appendix-B-Lemma-1-app}
	\text{Logm}(\mathbf{V}\tran\mathbf{A}\mathbf{V}) = \mathbf{V}\tran\text{Logm}(\mathbf{A})\mathbf{V} \\ \text{, iff }  \mathbf{V}\tran \mathbf{V} =  \mathbf{I} \text{ and $\mathbf{A}$ is SPD.}  
\end{aligned}
\end{equation}

\textbf{\textit{Proof:}} Based on the SPD property of matrix $\mathbf{A}$, we can apply ED to $\mathbf{A}$ and obtain:
\begin{equation}
\begin{aligned}
\label{Eq:appendix-B-Lemma-2}
\mathbf{A}\text{ is SPD }\Rightarrow \mathbf{A} &= \mathbf{V}^{A} \mathbf{D}^A \left( \mathbf{V}^{A}\right)\tran  \\ &\text{, where } \left( \mathbf{V}^{A}\right)\left( \mathbf{V}^{A}\right)\tran  =  \mathbf{I}
\end{aligned}
\end{equation}

Moreover, combining above results with $ \mathbf{V}\tran \mathbf{V} = \mathbf{I}$, we have

\begin{equation}
\begin{aligned}
\label{Eq:appendix-B-Lemma-3}
 \mathbf{V}\tran \mathbf{V}  =  \mathbf{V}\tran \left( \mathbf{V}^{A}\right) \left( \mathbf{V}^{A}\right)\tran  \mathbf{V} =  \mathbf{I} \\
 \Leftrightarrow  \left(  \mathbf{V}\tran \mathbf{V}^{A}  \right) \left( \mathbf{V}\tran \mathbf{V}^{A}   \right)\tran  =  \mathbf{I}
\end{aligned}
\end{equation}

Therefore, by substituting the results from equation (\ref{Eq:appendix-B-Lemma-2}) and (\ref{Eq:appendix-B-Lemma-3}) into $	\text{Logm}(\mathbf{V}\tran \mathbf{A}\mathbf{V})$, we have:
\begin{equation}
\begin{aligned}
\label{Eq:appendix-B-Lemma-4}
\text{Logm}(\mathbf{V}\tran\mathbf{A}\mathbf{V}) 
&\overset{\text{Eq. (\ref{Eq:appendix-B-Lemma-2})}}{=} \text{Logm}(\mathbf{V}\tran \mathbf{V}^{A} \mathbf{D}^A \left( \mathbf{V}^{A}\right)\tran \mathbf{V}) \\
&= \text{Logm}( \left(  \mathbf{V}\tran \mathbf{V}^{A}  \right) \mathbf{D}^A  \left( \mathbf{V}\tran \mathbf{V}^{A}   \right)\tran ) \\
&\overset{\text{Eq. (\ref{Eq:appendix-B-Lemma-3})}}{=} \left(  \mathbf{V}\tran \mathbf{V}^{A}  \right) \text{Logm}(  \mathbf{D}^A  ) \left( \mathbf{V}\tran \mathbf{V}^{A}   \right)\tran \\
&=  \mathbf{V}\tran\left[  \mathbf{V}^{A}   \text{Logm}(  \mathbf{D}^A  )\left( \mathbf{V}^{A}\right)\tran \right]  \mathbf{V} \\
&\overset{\text{Eq. (\ref{Eq:appendix-B-Lemma-2})}}{=}  \mathbf{V}\tran    \text{Logm}( \mathbf{V}^{A}   \mathbf{D}^A  \left( \mathbf{V}^{A}\right)\tran)  \mathbf{V} \\
&\overset{\text{Eq. (\ref{Eq:appendix-B-Lemma-2})}}{=}  \mathbf{V}\tran    \text{Logm}(\mathbf{A})  \mathbf{V} \\
\end{aligned}
\end{equation}

\textbf{Q.E.D.}

\section{Classification accuracy w.r.t. the number of applied filters}

\subsection{Classification accuracy w.r.t. the number of applied filters for all rest six datasets} \label{all_ds_acc}
\begin{figure*}[!htb]
	\centering
	\begin{subfigure}{.48\linewidth}
		\includegraphics[width=\linewidth]{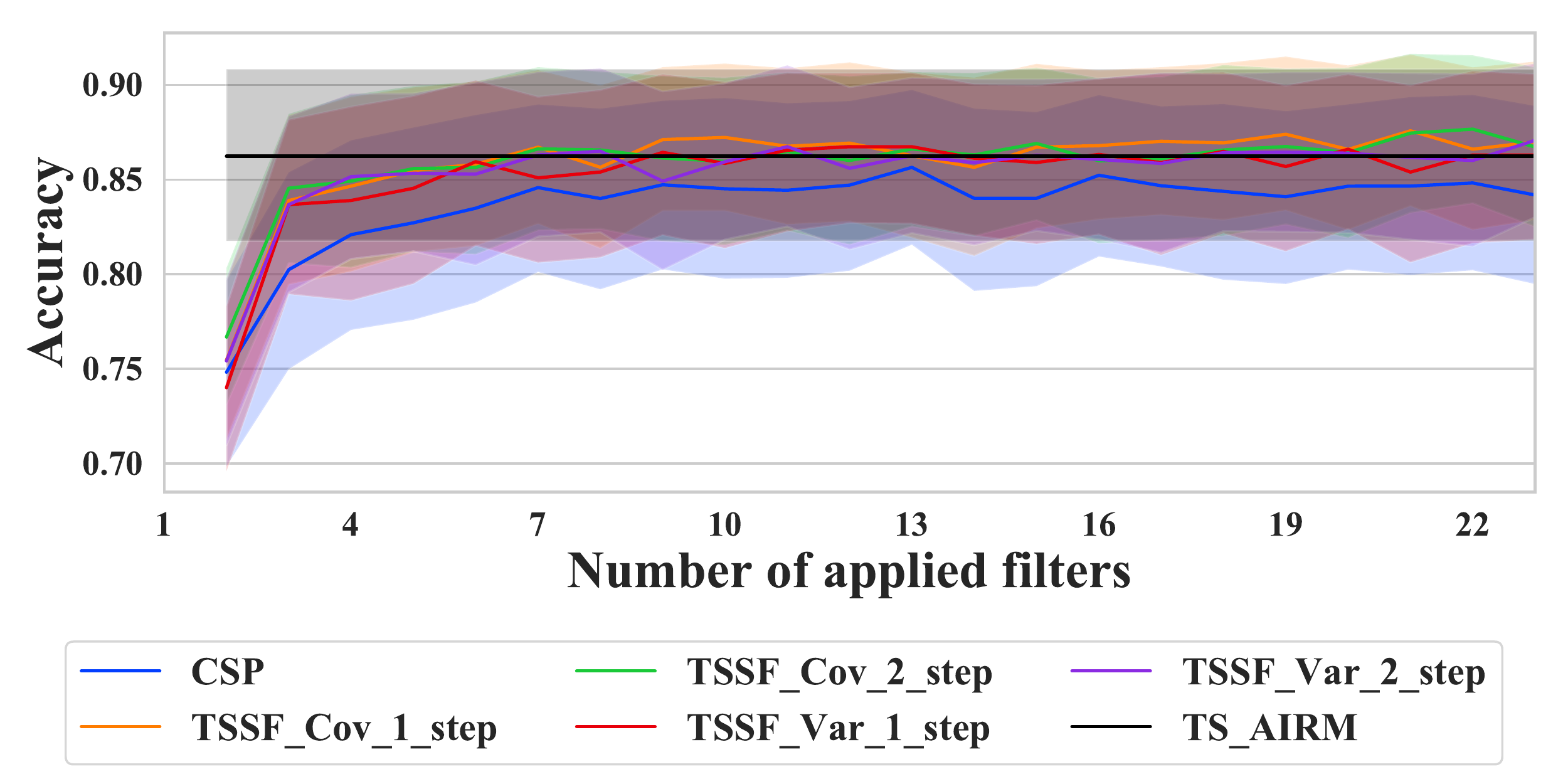}
		\caption{\centering BNCI2014001}
	\end{subfigure}
	\begin{subfigure}{.48\linewidth}
		\includegraphics[width=\linewidth]{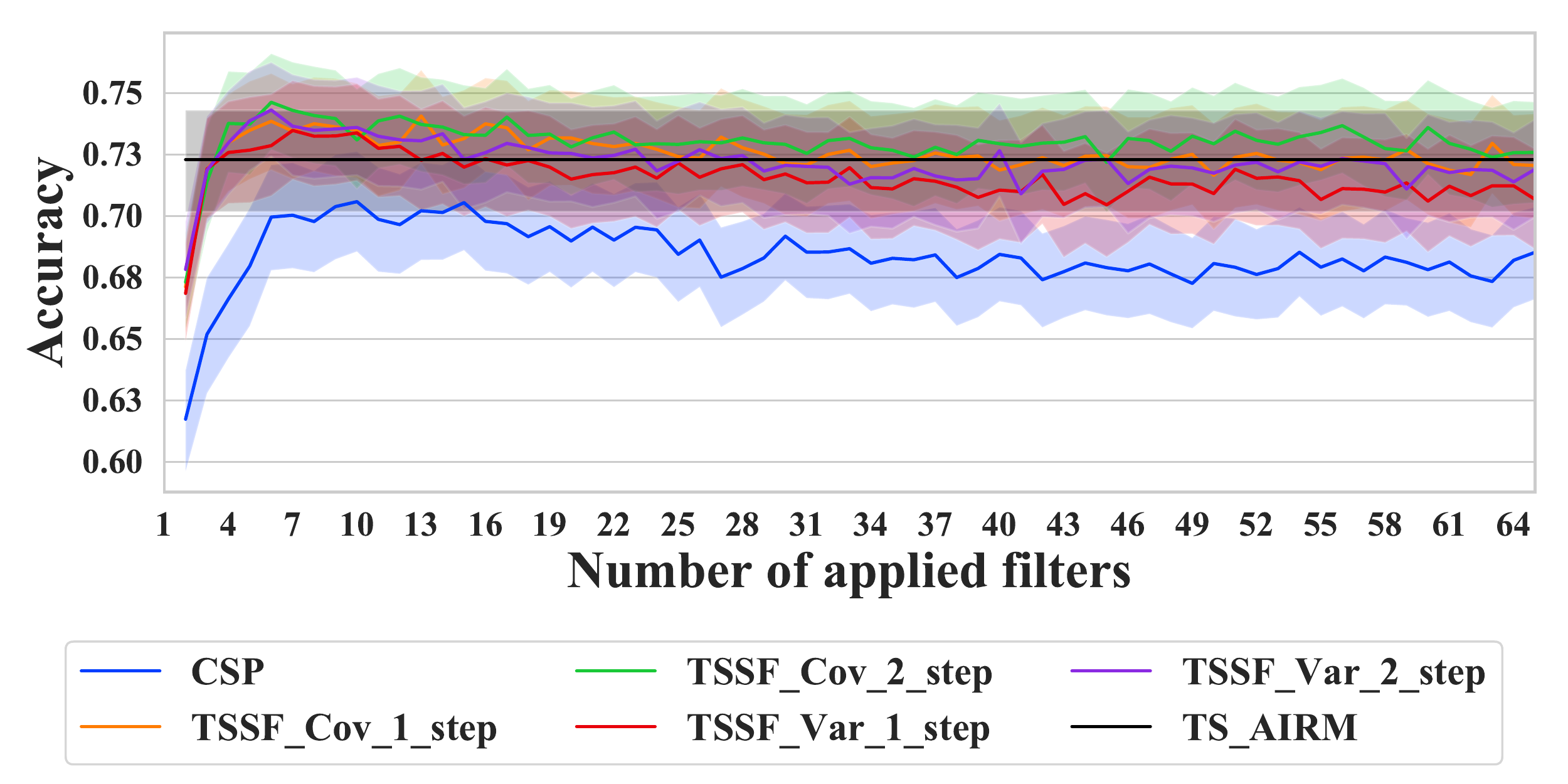}
		\caption{\centering Cho}
	\end{subfigure}
	\begin{subfigure}{.48\linewidth}
		\includegraphics[width=\linewidth]{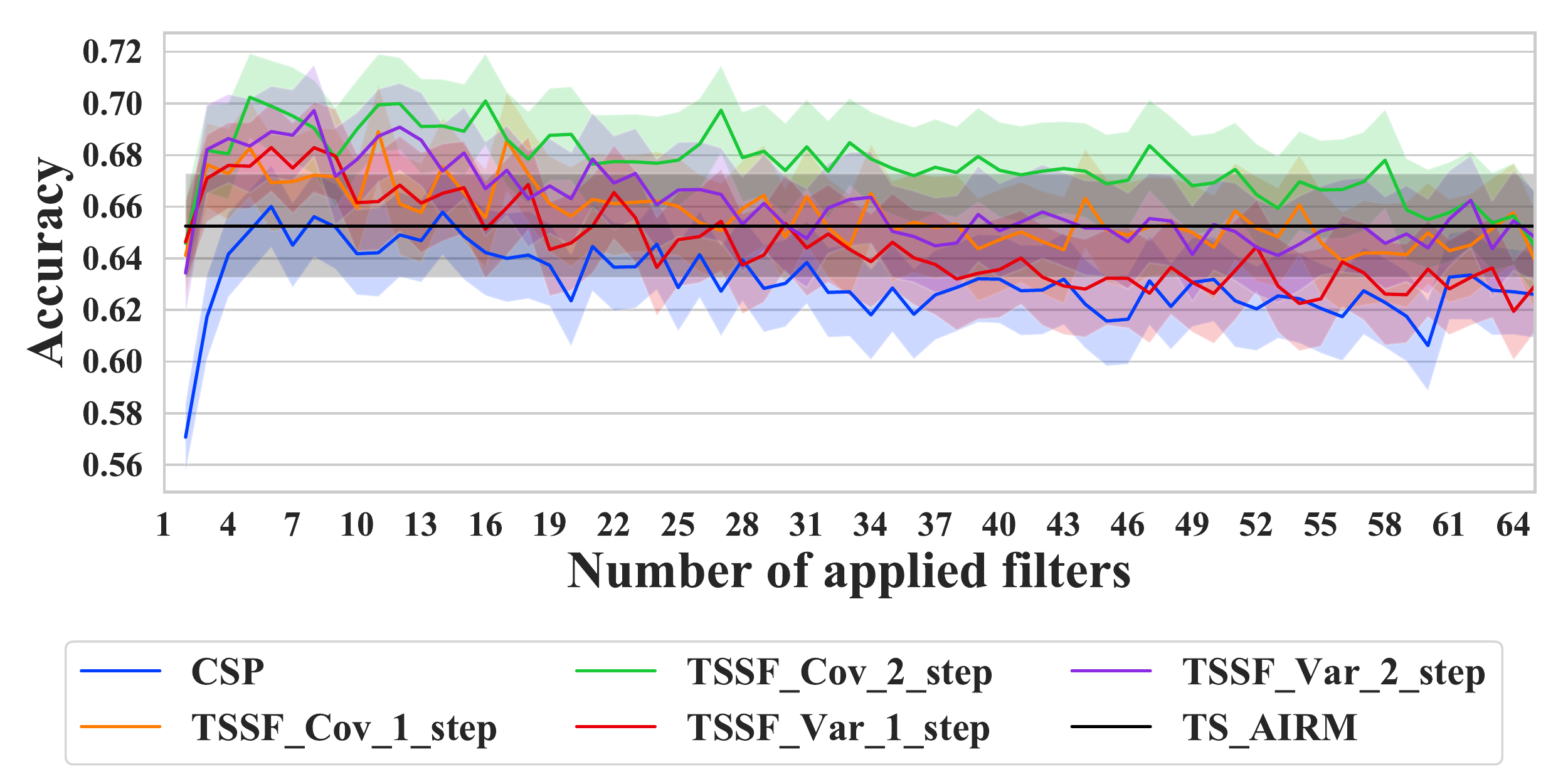}
		\caption{\centering Phy}
	\end{subfigure}
	\begin{subfigure}{.48\linewidth}
		\includegraphics[width=\linewidth]{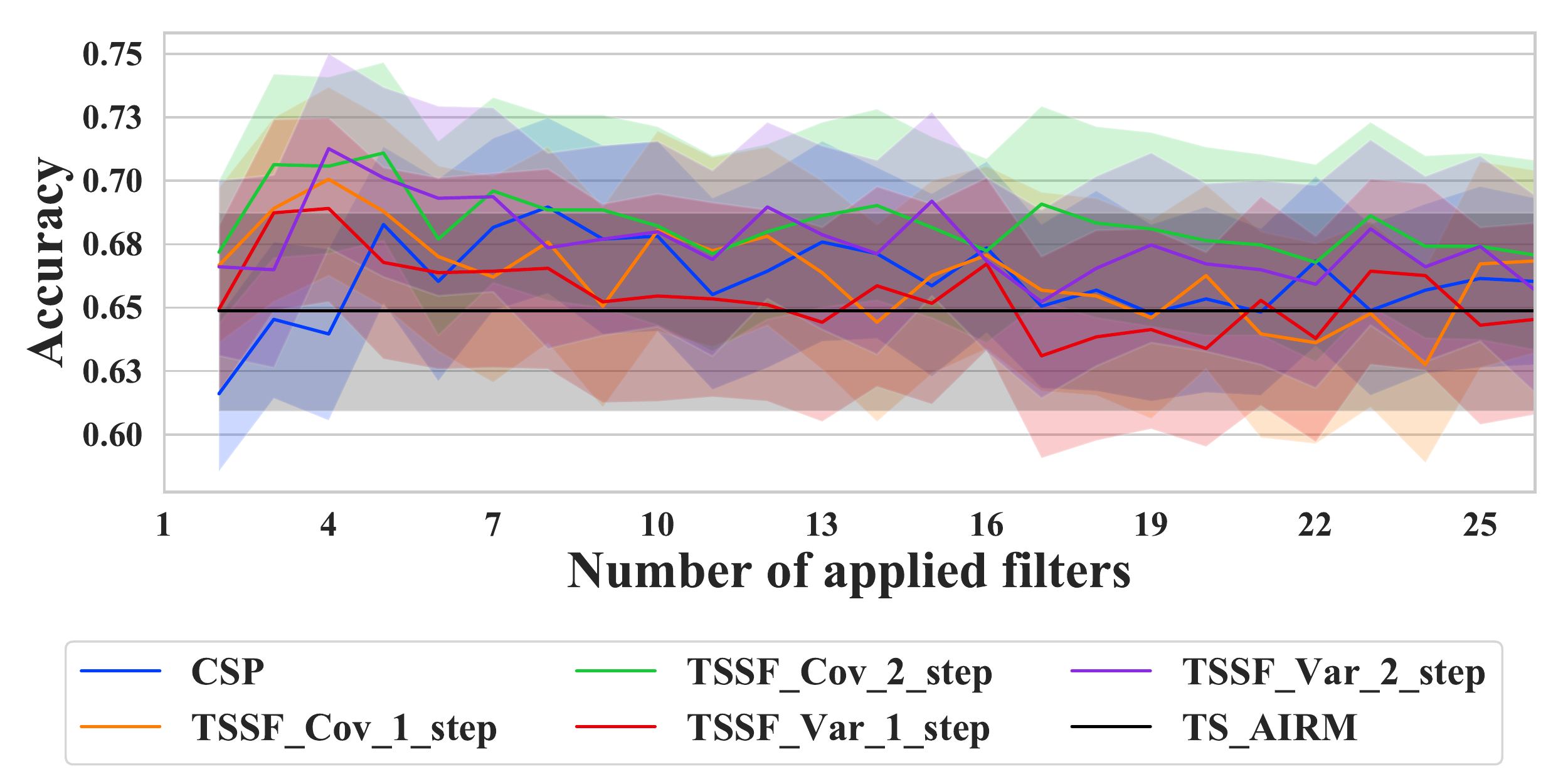}
		\caption{\centering Shin}
	\end{subfigure}
	\begin{subfigure}{.48\linewidth}
		\includegraphics[width=\linewidth]{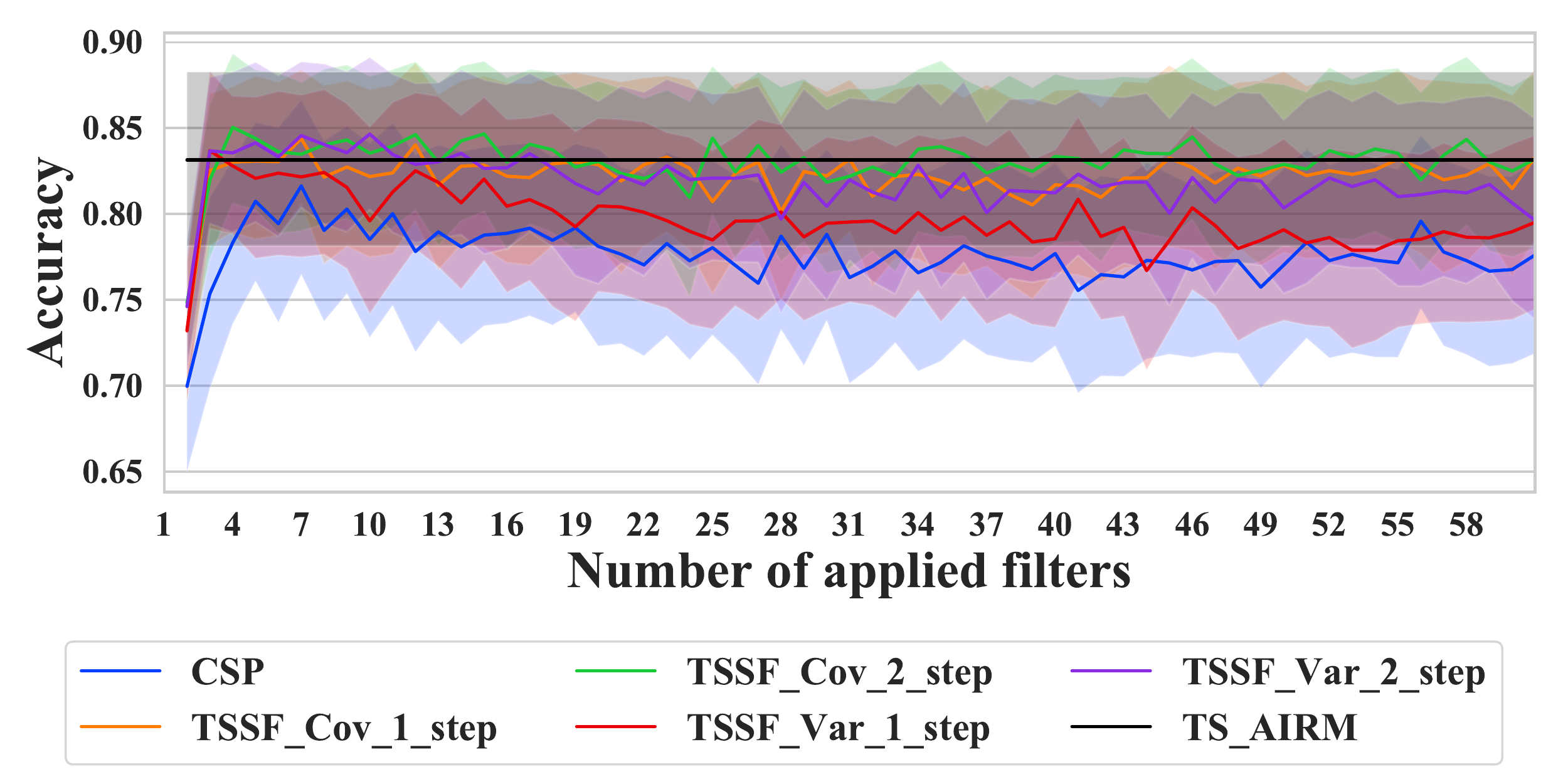}
		\caption{\centering Weibo}
	\end{subfigure}
	\begin{subfigure}{.48\linewidth}
		\includegraphics[width=\linewidth]{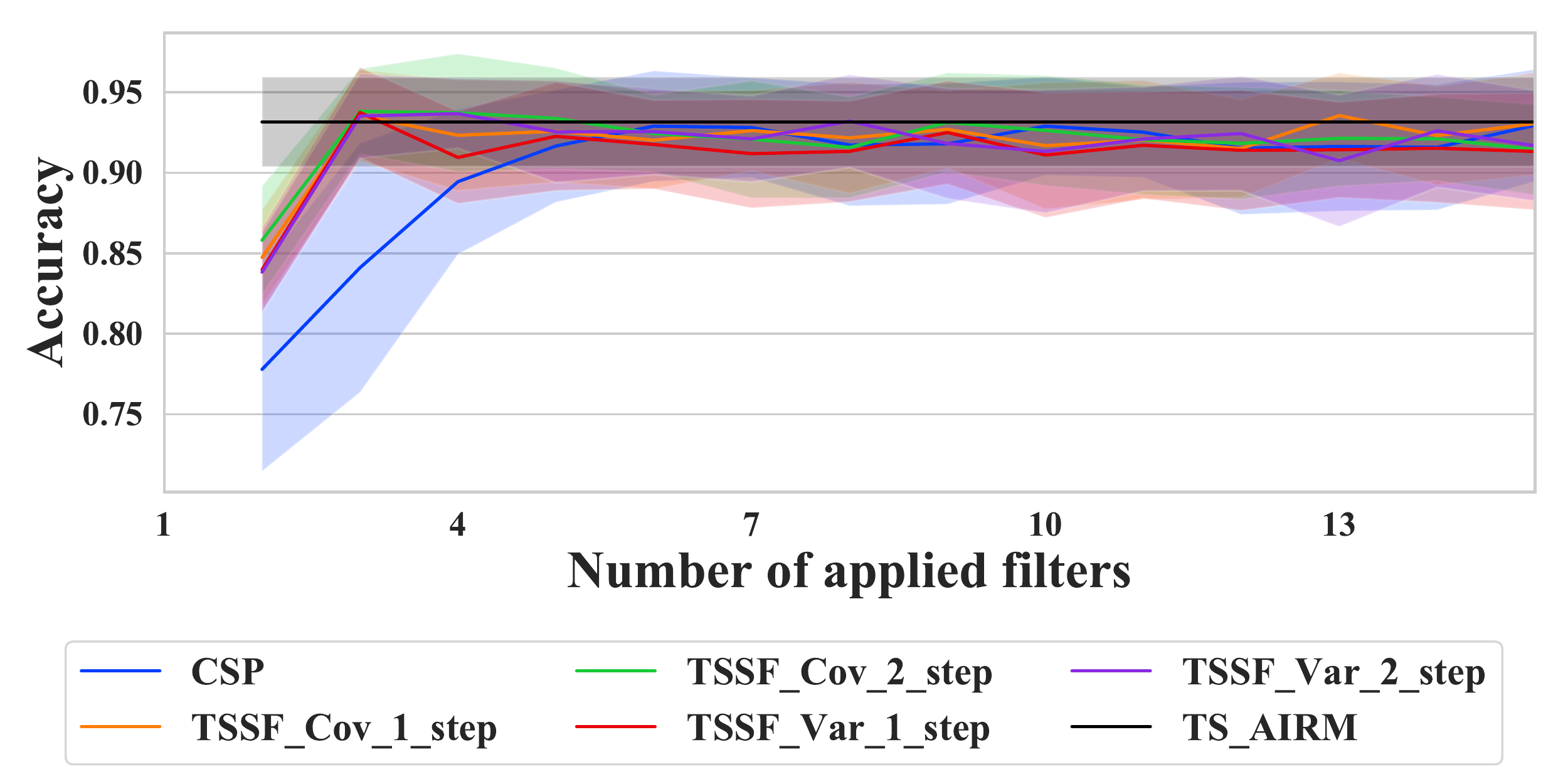}
		\caption{\centering Zhou}
	\end{subfigure}
	\caption{Classification accuracy w.r.t. the number of applied filters within each data set and the accuracies are computed across all subjects and sessions. The central line is the mean accuracy and the error band shows confidence interval = $68\%$.}
	\label{Fig_acc_allds}
\end{figure*}

(Please refer to Fig.~\ref{Fig_acc_allds})

\subsection{Classification accuracy w.r.t. the number of applied filters for all subjects within data set \textit{Munich Motor Imagery}} \label{Munich_acc_ind}
\begin{figure*}[!htb]
	\centering
	\includegraphics[width=\linewidth]{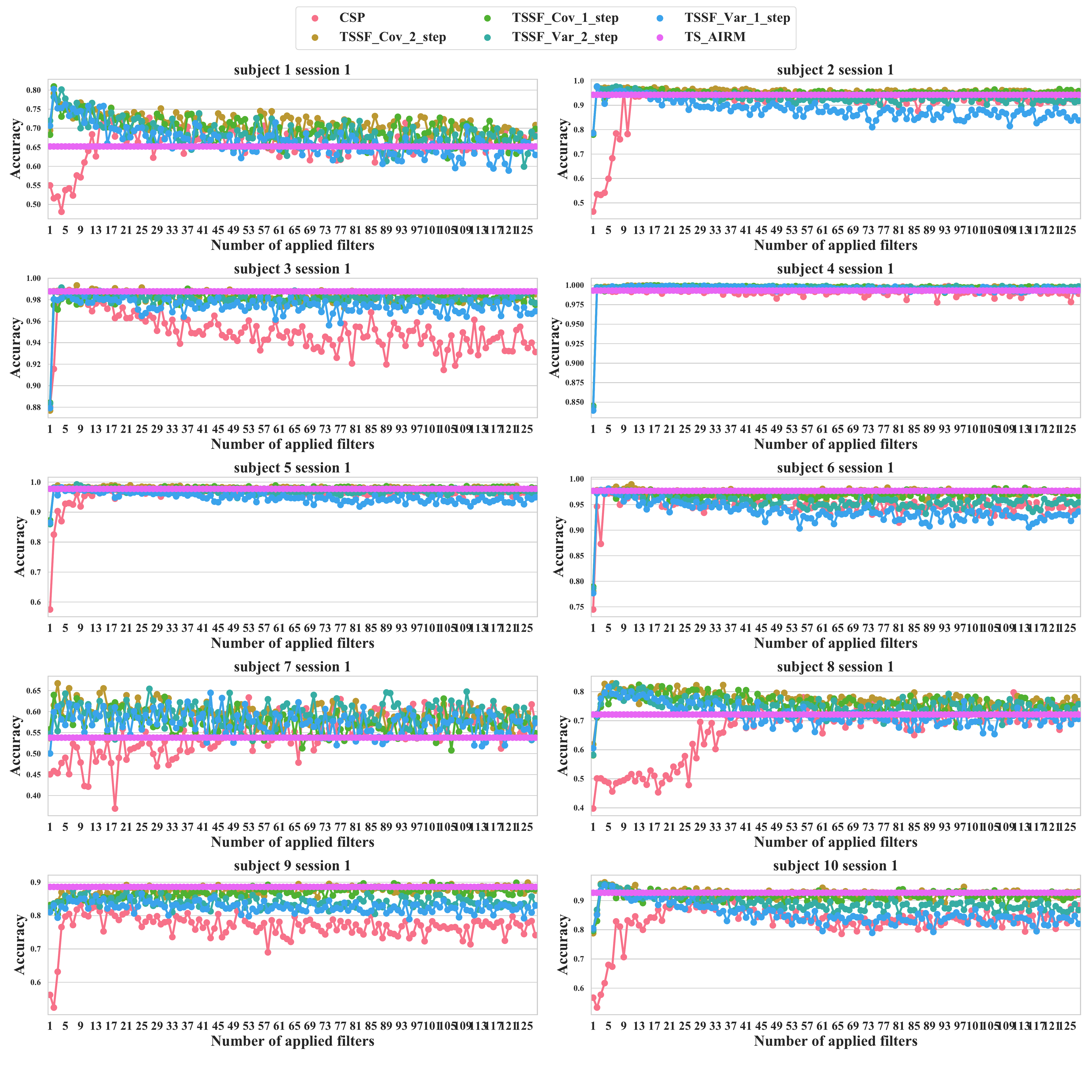}
	\caption{Classification accuracy w.r.t. the number of applied filters for each subject of data set \textit{Munich Motor Imagery} and the accuracies are computed using ROC-AUC metric and averaged across five folds cross validation.}
	\label{Fig_acc_ind}
\end{figure*}

(Please refer to Fig.~\ref{Fig_acc_ind})
\end{appendices}

\section*{Acknowledgment}

We would like to thank Alexandre Barachant for his input in conceiving of this project.

\printbibliography

\end{document}